\documentclass[prd,twocolumn,superscriptaddress,floatfix,amsmath,amssymb]{revtex4-1}
\usepackage[english]{babel}
\usepackage{graphicx}
\usepackage{verbatim}
\usepackage{mathrsfs}
\usepackage{hyperref}	
\usepackage{cancel}
\usepackage{bm}
\usepackage{color} 

\newcommand{\ket}[1]{\left| {#1} \right\rangle}

\newcommand{\proj}[2]{\left| {#1} \right\rangle\!\left\langle {#2} \right|}
\newcommand{\ii}{\mathrm{i}}

\newcommand{\tr}{\operatorname{Tr}}

\newcommand{\cnm}[2]{\big[{#1},{#2}\big]}

\newcommand{\Tr}{\text{Tr}}
\newcommand{\op}{\hat}
\newcommand{\da}{_\textsc{a}}
\newcommand{\db}{_\textsc{b}}

\begin{document}
\title{Causality issues of particle detector models in QFT and Quantum Optics}
\author{Eduardo Mart\'{i}n-Mart\'{i}nez}
\affiliation{Institute for Quantum Computing, University of Waterloo, Waterloo, Ontario, N2L 3G1, Canada}
\affiliation{Department of Applied Mathematics, University of Waterloo, Waterloo, Ontario, N2L 3G1, Canada}
\affiliation{Perimeter Institute for Theoretical Physics, Waterloo, Ontario, N2L 2Y5, Canada}

\begin{abstract}
We analyze the constraints that causality imposes on some of the particle detector models employed in quantum field theory in general and, in particular, on those used in quantum optics (or superconducting circuits) to model atoms interacting with light. Namely, we show that disallowing faster-than-light communication can impose severe constraints on the applicability of particle detector models in three different  common scenarios: 1) when the detectors are spatially smeared, 2) when a UV cutoff  is introduced in the theory and 3) under one of the most typical approximations made in quantum optics: the rotating-wave approximation. We identify the scenarios in which the models' causal behaviour can and cannot be cured. 
\end{abstract}

\maketitle

\section{Introduction}

Particle detector models may be thought of as controllable quantum systems that couple locally in space and time to quantum fields.  They provide a way to extract localized spatio-temporal information from the fields without having to implement  projective measurements of  localized field observables \cite{Drago1,Drago2}. This simplifies the task of extracting localized information about the field, and avoids all the possible complications that may appear with the use of projective measurements altogether \cite{Lin2014773}. Particle detector models in quantum field theory  were pioneered by Unruh and DeWitt \cite{unruh_notes_1976,DeWitts}, and can be found in the literature in several slightly different (but fundamentally similar) formats, e.g., a field in a box \cite{unruh_notes_1976}, a two-level system \cite{DeWitts} or a harmonic oscillator (see, e.g., \cite{BeiLok,Brown2012,Fuenetesevolution}).

Particle detector models have been successfully employed  in a plethora of contexts in fundamental quantum filed theory \cite{Crispino,Takagi}. Perhaps one of the best-known ones is the operational formulation of the Hawking and Unruh effects (see, e.g., \cite{unruh_notes_1976,candelas_irreversible_1977}). Besides their many uses in fundamental quantum field theory, particle detector models are ubiquitous as models for experimental setups in quantum optics \cite{scullybook} and in superconducting circuits \cite{Wallraff:2004aa}. For example, an alkali atom as a first quantized system, can serve as such a detector for the second quantized electromagnetic field. In fact, the common light-matter interaction models, such as for instance the Jaynes-Cummings model and its variants, are almost identical in nature to the Unruh-DeWitt (UDW) model \cite{scullybook}. Indeed, the Unruh-DeWitt detector-field interaction has been proven to be a good model of the light-matter interaction in quantum optics for processes not involving  exchange of orbital angular momentum \cite{Wavepackets,Alvaro}.

More recently, UDW detectors have been extensively used  in  studies of  relativistic quantum information. For example, relativistic quantum computing \cite{AasenPRL,Chris}, quantum communication via field quanta \cite{mathieuachim1,Robort,Robort2}, cosmology \cite{Gibbons1977,QuanG,cosmoq}, and, more generally, in studies on a host of effects related to the presence of spacelike  entanglement in the vacuum state of quantum fields both from fundamental \cite{Valentini1991,Reznik2003,Reznik1,VerSteeg2009,Olson2011,Salton:2014jaa,Pozas-Kerstjens:2015,Martin-MartinezSmithTerno} and applied \cite{Farming,Brown:2014en} perspectives. Interestingly, in these studies it is shown that it is possible to harvest correlations from the field vacuum to spacelike separated detectors, which gives an operational proof of the spacelike entanglement present in the quantum vacuum \cite{Alegbra1,Alegbra2}.

Because UDW detector models have been proven so valuable in relativistic quantum information, the question arises as to what extent these models (which involve non-relativistic systems coupled to fully relativistic quantum fields) behave in a causal way. This is of special importance when studying phenomena for which the causal behaviour of the model is paramount, such as the aforementioned spacelike vacuum entanglement harvesting or, more generally, any quantum communication scenario where relativistic effects become important.

 It is known that naive projective measurements in quantum fields are not compatible with causality and can suffer from superluminal signalling (see for instance \cite{Lin2014773}). Do UDW-like detector models have similar problems? In particular, in \cite{BenicasaBorstenBuckDowker} this same question is posed about the causal behaviour of the Unruh-Dewitt model. The question is well aimed: consider, for example, the usual oversimplified single-mode Jaynes-Cummings model \cite{scullybook}. This model allows for faster-than-light signalling and is indeed unable to model the light matter interaction in relativistic regimes. The question whether the full UDW model respects causality was addressed in \cite{mathieuachim1}, where it was shown that communication using pointlike Unruh-DeWitt detectors as emitters and receivers of field quanta is causal to any order in perturbation theory, as long as the detectors remain pointlike as originally proposed.

However, as we will discuss, it is all too common to consider modifications of the original model that  sometimes jeopardize its causal behaviour, both in theoretical studies on field theory and in the applications of these models to quantum optics.  Studying the causal behaviour of these modified models has a double interest: on the one hand, from a theoretician's point of view, in the scenarios where the model presents causality problems it is interesting to know if and when the causal behaviour can be restored in an approximate way in some regimes. On the other hand, as an experimentalist using these models to predict phenomenology, it is fundamental to know under what circumstances the models can be used to make meaningful predictions and, conversely, when the models dramatically fail due to violations of a first principle of relativistic theory. 

The first modification of the original model whose causal behaviour we are going to study is the generalization to non-pointlike detectors. It is indeed very common to consider  that the detectors have a spatial smearing \cite{Langlois2006,Satz2006}. The smearing of particle detectors may respond to the need to regularize divergences of the pointlike model \cite{Schlicht}. Smearings can also be introduced, for example in quantum optics, to improve on the accuracy of the light-matter interaction models assuming that the atoms are not pointlike objects, but are instead localized in the full extension of their atomic wavefunctions  \cite{scullybook,Alvaro}. Furthermore,  as discussed in \cite{Schlicht} oftentimes spatial smearings are introduced implicitly, hidden in some form of soft UV regularization of the model (see, e.g., \cite{Takagi}).

 As was already noted in \cite{BenicasaBorstenBuckDowker}, and as we will study in detail in this paper, spatially compactly supported detectors are safe in terms of their causal behaviour. However, the use of compactly supported spatial smearing is not so common in the literature. Instead, the most common smeared UDW models used in the literature (both in QFT and quantum optics) assign non-compact ---but very strongly supported in a finite region--- smearing functions, such as Gaussian (See, e.g. \cite{Pozas-Kerstjens:2015,Wavepackets})  or Lorentzian (See, e.g, \cite{Schlicht,Langlois2006,Satz2006}) smearings. Non-compact smearings are employed mainly for two reasons: 1) the simplification of analytic computations and the regularization of divergences and, 2) In the context of quantum optics, the introduction of smearing is part of the refinement of the model: the atomic wavefunctions of valence electrons characterizing the effective size of the atom are exponentially suppressed in the radial distance to the centre of mass of the atom \cite{Alvaro}. 
 
A relevant question to ask in these scenarios is to what extent the use of fast-decaying smearing functions (as opposed to strictly compactly supported ones) effectively renders the predictions of the model acausal. Rigorously speaking, any non-compactly supported smearing allows for instantaneous signalling. For example, if we consider communication between two non-compact detectors A and B, the influence of A in B would instantaneously be felt by B. Nevertheless, if in spite of being non-compact, the smearing of the two detectors is only strongly supported around a point of space with a characteristic length $\sigma$ (as is the case of Gaussian or Lorentzian smearing), causality may be recovered in an approximate sense when we consider, for instance, communication between two detectors whose spatial separation is much larger than $\sigma$. In this paper we will study quantitatively how the leading-order causality violations in the communication between two detectors are suppressed faster than exponentially as the characteristic size-scale of the non-compact smearing is reduced. Hence, this leads us to conclude that the causality of the model can indeed be approximately recovered with arbitrary precision when using strongly-supported but non-compact smearings such as Gaussian.

Another very typical modification of the UDW model, fairly common in quantum optics, is the introduction of hard UV cutoffs in the detector's proper reference frame. This is usually justified by the fact that the atomic response to electromagnetic radiation is a function of frequency and it does eventually become negligible in the far UV limit. If we were to consider such as an effective model, we would have to keep in mind that when a UV cutoff is introduced the causality of the model is again compromised. How causality is impacted by this sort of hard UV cutoffs has been studied in cavity settings both perturbatively \cite{Robort} and non-perturbatively \cite{Brown2012}. These studies shown that causality violating vanish decrease polynomially fast as the UV cutoff is relaxed. However, these studies had two limitations: they were limited to 1+1-dimensional setups and only to a discrete set of field modes corresponding to quantum fields in periodic and Dirichlet cavities. The fact that we consider a continuum of modes (field in free space) as opposed to a discrete number of them does introduce differences in the causal behaviour of the UDW when UV cutoffs are introduced. Much more important, as we will see in this work, the dimensionality of spacetime is critical for the causal behaviour of the communication using UDW detectors in the presence of UV cutoffs. For example, we will discuss that in a 3+1-dimensional scenario the causality loss due to the existence of these cutoffs cannot be made arbitrarily small just increasing the cutoff without further regularization, in stark contrast with what was obtained for 1+1-dimensional cavities in \cite{Brown2012,Robort}.  

Finally, we will analyze the impact of causality on the validity of the {\it rotating-wave approximation} (RWA). The RWA \cite{scullybook} is overwhelmingly present in the literature on quantum optics. The approximation consists of the simplification of the light-matter interaction Hamiltonian removing the terms that oscillate rapidly. The RWA is discussed to be valid for long interaction times (much larger than the detector's characteristic Heisenberg time). However, this approximation may result extremely harmful for the causality of the model. This issue has already drawn the attention of several researchers in early \cite{Tiramisu,Milo} and more recent \cite{Dolce,NonRWA} works. We will analyze the faster-than-light signalling that this approximation enables in the context of quantum communication of particle detectors through the exchange of field quanta. We will show that even for long interaction times, the acausal signature of this extremely common approximation is only very slowly erased as the limit of longer times is taken, rendering this approximated model unfit to describe any setting where the causality of the theoretical model is important.

With this perturbative study of the causal behaviour of the UDW model and the light-matter interaction, we will quantitatively characterize if (and in what regimes) the approximations and considerations described above can still be used mantaining some degree of approximate causal behaviour. This is important because current experimental techniques in quantum optics and superconducting circuits are nearing the point where intrinsically relativistic phenomena  can be accessed in experiments (see e.g. \cite{DCasimir}).  Finally, we note that we do a general study in one, two and three spatial dimensions with a double purpose: 1) There exist experimental setups where quantum fields live in an effectively reduced dimensionality such as superconducting microwave guides coupling to artificial atoms \cite{Wallraff:2004aa,supercond,Ultrastrong}, and  2) in doing so, we will be able to characterize the relevant role that the number of spatial dimensions has in the causal behaviour of the light-matter interaction models, and learn some interesting aspects of the regular UV behaviour of the signalling between two particle detectors.

\section{Signalling of two particle detectors through a quantum field}

We will study the Unruh-DeWitt particle detector (from now on referred to as the `detector') which will be considered in this work as a spatially localized two-level quantum system coupled to a scalar field.  The spatial smearing of the detector will be given by the real-valued smearing function $F(\bm x)$ which we choose, for convenience and w.l.g., to be localized and centred around $\bm x=0$.  This detector couples to a scalar field locally along its trajectory in spacetime. In general, we can  consider that the  detector's centre of mass moves in a trajectory parametrized in terms of its proper time $\tau$. That is, $\bm x=\bm x_0(\tau)$. In the particular case of a stationary detector whose centre of mass is placed at $\bm x=\bm x_0$, the detector is comoving with the usual Minkowski frame $(\bm x, t)$ in which we carry out the field quantization. This means that we can take $t=\tau$. The Unruh-DeWitt  interaction Hamiltonian (in the interaction picture) for this case is given by \cite{Langlois2006}
\begin{equation}\label{puno}\op H_I=\lambda \int \text{d}^n \bm x\, F[\bm x-\bm x_0] \chi(t)\op m(t)\op \phi[\bm x,t]\end{equation}
where  $\chi(t)$ is the switching function controlling the coupling-decoupling speed and the duration of the detector-field interaction, \mbox{$\op m(t) =(\op\sigma^+e^{\ii\Omega t}+\op\sigma^-e^{-\ii\Omega t})$} is the detector's monopole moment (being $\Omega$ the energy gap between the detector's two energy levels), $ \op\phi[\bm x,t]$ is the quantum scalar field, and $n$ is the number of spatial dimensions. The difference of \eqref{puno} with the pointlike model is that the field operator is evaluated along the worldline of the detector, as usual, but summed over the whole spatial extension of the detector. 

Although this is out of the scope of the present paper, it is worth mentioning that if we were to generalize the analysis of these localized detector models to general non-inertial trajectories, we would have to face the well-known problem of accelerating rigid
bodies. Roughly speaking,  the proper distance between two points of a solid accelerating  with  the  same  proper acceleration  increases in time, eventually destroying the solid when the  internal  cohesion forces  that  support it  are  overrun by ``stress forces'' \cite{BellParadox}.  A  reasonable  hypothesis  made for  a  physical  detector in those cases  is that, until this happens, the detector has to keep internal coherence to a good approximation.  In other words, the  internal  forces  that  keep  the  detector  together  will prevent it from being further smeared due to relativistic effects, up to some threshold acceleration.  This in turn implies that,  effectively,  every point of the detector will accelerate with a different acceleration in order to keep up with the rest of its points and maintain the shape of the spatial profile of the detector in the centre-of-mass reference frame.  The natural framework to treat  such a  detector  is  the  use  of  the  well-known  Fermi-Walker frame. If smeared detectors undergoing non-inertial trajectories are considered approximately rigid and sustained by internal cohesion forces, the smearing function becomes a function of the Fermi-Walker coordinates on the spacelike orthogonal hypersurface to the detector's motion.  How this localization works out for general trajectories of the detector is very well detailed in \cite{Schlicht} (or among other references, in, for example, \cite{Langlois,Satz2006,Wavepackets}). 

Back to our original scenario, consider now two stationary Unruh-DeWitt detectors A and B whose profiles are centred at positions $\bm x_\textsc{a}$ and $\bm x_\textsc{b}$. They are mutually at rest and they share a proper time $t$. The interaction picture Hamiltonian describing the interaction of these two detectors with the field is
\begin{align}
 \op H_I&=\sum_{\nu}\lambda_\nu  \chi_\nu(t)\op\mu_\nu(t)\int \text{d}^n\bm x\ F(\bm x-\bm x_\nu)\op\phi(\bm x, t),
\end{align}
where   $\nu=\{\text{A},\text{B}\}$ labels the detectors, and the monopole moment operators $\op\mu_\nu(t)$ act on the two-detector Hilbert space $\mathcal{H}_\textsc{a}\otimes\mathcal{H}_\textsc{b}$ according to
\begin{equation}\label{poddard}
\op\mu\da(t)\equiv\op m\da(t)\otimes \openone\db,\qquad \op\mu\db(t)\equiv \openone\da\otimes \op m\db(t),
\end{equation}
where
\begin{equation}
\op m_\nu (t) =(\op\sigma_\nu^+e^{\ii\Omega_\nu t}+\op\sigma_\nu^-e^{-\ii\Omega_\nu t}).
\end{equation}

Let us now consider an arbitrary initial state density matrix  $\op\rho_0$ for the system comprised of the two detectors and the field. After the interaction of the detectors with the field, modulated by the switching functions $\chi_\nu(t)$, the time-evolved state will be given by
$\op\rho_T = \op U \op\rho_0 \op U^\dagger$, 
where $\op U$ is the interaction picture time evolution operator. 
Assuming that each $\lambda_\nu$ is small enough to be in the perturbative regime , we can consider the Dyson expansion of the time-evolution operator
$\op U$:
$\op U =\op  U^{(0)} + \op U^{(1)} +\op  U^{(2)} 
+ \mathcal{O}(\lambda_\nu^3)$ 
where $\op U^{(0)} = \openone$ 
and
\begin{align} 
\op U^{(1)}\!\!
=\! 
- \ii\!\!\int_{-\infty}^{\infty}\!\!\!\!\!\! \text{d} t 
\, H(t)
, \;
\quad\op U^{(2)}\! =\! 
- \!\!\int_{-\infty}^{\infty}
\!\!\!\!\!\text{d}t \!\int_{-\infty}^{t}\!\!\!\!\!\!\text{d}t'
\op H(t) \op H(t')
\ . 
\label{eq:U2-def}
\end{align}
Hence 
$\op\rho_T  = \op\rho_0+\op\rho_T^{(1)}+\op\rho_T^{(2)}+\mathcal{O}(\lambda^3)$, where 
\begin{align}
\label{eq:rho1}\op\rho_T^{(1)}
&=
U^{(1)}\op\rho_0+\op\rho_0{U^{(1)}}^\dagger 
\ , 
\\
\label{eq:rho2}\op\rho_T^{(2)}
&=
U^{(1)}\op\rho_0{U^{(1)}}^\dagger+U^{(2)}\op\rho_0+\op\rho_0{U^{(2)}}^\dagger
\ . 
\end{align} 
The final state of the two-detector subsystem will, hence, be given by
\begin{align}
\label{rhoexp}
\op\rho_{\text{d},T} &= \Tr_\phi(\op\rho_T)
= 
\op\rho_{\text{d},0} +\op\rho_{\text{d},T}^{(1)}+ \op\rho_{\text{d},T}^{(2)}
+ 
\mathcal{O}(\lambda^3) 
\ , 
\end{align}
where $\tr_\phi$ denotes the trace over the field Hilbert space. 

As we will detail below, we are going to consider that detector A interacts with the field earlier in time, and detector B will couple to the field afterwards. We would like to analyze the ability of A to signal B. In particular, we would like to see what is the influence of the existence of A in the time evolved quantum state of B. Obviously, in a causal model, detector B's state cannot depend in any way on the initial state of detector A if A and B remain spacelike separated during their interaction with the field. If we assume the most general uncorrelated initial state for the detectors and the field, the initial state density matrix takes the general form 
\mbox{$\op\rho_0=\op\rho_{\text{d},0}\otimes \op\rho_{\op\phi,0}$}, 
where $\op\rho_{\text{d},0}$ and $\op\rho_{\op\phi,0}$ 
are respectively the initial state of the 
two-detector subsystem and the initial state of the field.  The leading-order  contributions of the influence of detector A on detector B will be proportional to $\lambda_\textsc{a}\lambda_{\textsc{b}}$. The first order term in \eqref{rhoexp} hence does not contribute to the signalling between A and B. Let us then focus on the second order contribution. Under the assumption that $\op\rho_0=\op\rho_{\text{d},0}\otimes \op\rho_{\op\phi,0}$, the second order contribution to the time evolved state of the two detectors will be given by  
\begin{align}
\op\rho_{\text{d},T}^{(2)}
& =\sum_{\nu,\eta}\lambda_\nu\lambda_\eta\bigg[ \int_{-\infty}^{\infty}\!\!\!\text{d}t \int_{-\infty}^{\infty}\!\!\!\text{d}t' \, 
\chi_\nu(t')\chi_\eta(t) 
\notag
\\
&\hspace{13ex}
\times \op\mu_\nu(t') \op\rho_{\text{d},0} \op\mu_\eta(t) \, W[\bm x_\eta,t,\bm x_\nu,t']
\notag
\\
&\hspace{4ex}
- \int_{-\infty}^{\infty}\!\!\!\text{d}t \int_{-\infty}^{t}\!\!\!\text{d}t' \, 
\chi_\nu(t)\chi_\eta(t') 
\notag
\\
&\hspace{13ex}
\times \op\mu_\nu(t)\op \mu_\eta(t') \op\rho_{\text{d},0} \, 
W[\bm x_\nu,t,\bm x_\eta,t']
\notag
\\
&\hspace{4ex}
- \int_{-\infty}^{\infty}\!\!\!\text{d}t \int_{-\infty}^{t}\!\!\!\text{d}t' \, 
\chi_\nu(t)\chi_\eta(t') 
\notag
\\
&\hspace{13ex}
\times\op\rho_{\text{d},0}\op \mu_\eta(t') \op \mu_\nu(t) \, 
W[\bm x_\eta,t',\bm x_\nu,t]\bigg] 
\label{eq:rho-d-T-2}
\end{align}
Where $W[\bm x_\nu,t,\bm x_\eta,t']$ denotes 
the pullback of the Wightman function on the detectors' smeared worldlines, 
\begin{align}\label{whightman1}
 W[\bm x_\nu,t,\bm x_\eta,t'] &=\!\!\int\! \text{d}^n\bm x\!\!\int\! \text{d}^n\bm x' F[\bm x\!-\!\bm x_\nu]\\ 
\nonumber&\!\!\!\!\!\!\!\!\!\!\!\!\!\times F[\bm x'\!-\!\bm x_\eta]\Tr_{\op\phi}
\big[ \op \phi\bigl({\bm x},t\bigr)\op \phi\bigl({\bm x'},t'\bigr)\op\rho_{\op\phi,0}\big].
\end{align}
Again $\Tr_{\op\phi}$ denotes the partial trace over the field Hilbert space.


Since we are going to analyze the signalling between detectors A and B, we will not be interested in the local terms in \eqref{eq:rho-d-T-2}, that is, the terms that would vanish if one of the detectors is not present. With this in mind,  we can decompose \eqref{eq:rho-d-T-2} as
\begin{equation}\label{eqnoissig}
\op\rho_{\text{d},T}^{(2)}=\lambda_\textsc{a}\lambda_\textsc{b}\op\rho_{\text{signal}}^{(2)}+\sum_\nu\lambda_\nu^2\op\rho_{\nu,\text{noise}}^{(2)}
\end{equation}
where the terms $\op\rho_{\nu,\text{noise}}^{(2)}$ contain only contributions proportional to $\lambda_\nu^2$, and as such they are local to each detector and they are not involved in the flow of information from A to B. One quick way to see this is to notice that in the partial state of detector B, i.e. $\op\rho_{\textsc{b},T}=\tr_\textsc{a} \op \rho_{\text{d},T}$, the only leading-order contributions that depend on the existence of detector A at all are proportional to $\lambda_\textsc{a}\lambda_\textsc{b}$.

The terms $\op\rho_{\text{signal}}^{(2)}$  can be further simplified assuming that the detector's switching functions are compactly supported and that their supports do not overlap in time. Without loss of generality let us assume that the detector A is switched on before B, and then let us make the additional assumption that
\begin{align}\label{condos}
\nonumber \text{supp}[\chi_\textsc{a}(t)]=[T_\textsc{a}^{\text{on}},T_\textsc{a}^{\text{off}}],\quad\text{supp}&[\chi_\textsc{b}(t)]=[T_\textsc{b}^{\text{on}},T_\textsc{b}^{\text{off}}],\\
T_\textsc{b}^{\text{on}}>T_\textsc{a}^{\text{off}}.
\end{align}
Imposing this assumption on \eqref{eq:rho-d-T-2}, we can simplify $\op\rho_{\text{signal}}^{(2)}$ to
\begin{align}\label{towardsit}
&\op\rho_{\text{signal}}^{(2)}= \int_{-\infty}^{\infty}\!\!\!\text{d}t \int_{-\infty}^{\infty}\!\!\!\text{d}t' \, \chi_\textsc{a}(t)\chi_\textsc{b}(t') \\
\nonumber&\!\!\times\!\!\bigg[W[\bm x_\textsc{b},t',\bm x_\textsc{a},t] \Big(\op\mu_\textsc{a}(t)\op\rho_{\text{d},0}\op \mu_\textsc{b}(t')\!-\! \op\mu_\textsc{b}(t')\op\mu_\textsc{a}(t)\op\rho_{\text{d},0}\Big)\\
\nonumber&\!\!+\!W[\bm x_\textsc{a},t,\bm x_\textsc{b},t'] \Big(\op\mu_\textsc{b}(t')\op\rho_{\text{d},0}\op \mu_\textsc{a}(t)\!-\! \op\rho_{\text{d},0} \op\mu_\textsc{a}(t)\op\mu_\textsc{b}(t')\Big)\bigg]\!. 
\end{align}

In a signalling scenario under the assumption \eqref{condos}, since A interacted with the field first, we can regard A as the sender and B as the receiver. All the information transmitted through the field from A to B will be encoded (to leading order) in the parts of the density matrix of B that are proportional to $\lambda\da\lambda\db$. This signalling contribution to B's partial state will be given by
\begin{equation}\label{porw}
\op\rho_{\textsc{b},\text{signal}}^{(2)}=\tr\da(\op\rho_{\text{signal}}^{(2)}).
\end{equation}

Recalling \eqref{poddard} and making the reasonable assumption that prior to communication the initial state of the two detectors is uncorrelated (i.e., \mbox{$\op\rho_{\text{d},0}=\op\rho_{\textsc{a}}\otimes \op\rho_{\textsc{b}}$}),  we can compute the explicit form of \eqref{porw} from \eqref{towardsit}:
\begin{align}\label{traceda}
&\op\rho_{\textsc{b},\text{signal}}^{(2)}= \int_{-\infty}^{\infty}\!\!\!\text{d}t \int_{-\infty}^{\infty}\!\!\!\text{d}t' \, \chi_\textsc{a}(t)\chi_\textsc{b}(t')\tr[\op m_\textsc{a}(t)\op\rho_{\textsc{a}}] \\
\nonumber&\qquad\times 2\ii\,\text{Im}\bigg[ W[\bm x_\textsc{a},t,\bm x_\textsc{b},t'] \bigg] \Big(\op m_\textsc{b}(t')\op\rho_{\textsc{b}}\!-\! \op\rho_{\textsc{b}}\op m_\textsc{b}(t')\Big)
\end{align}
where we have also used that
\begin{equation}
W[\bm x_\textsc{a},t,\bm x_\textsc{b},t']=\Big(W[\bm x_\textsc{b},t',\bm x_\textsc{a},t]\Big)^*
\end{equation}
which is very easy to check form \eqref{whightman1}.

For each of the the individual detectors we may introduce 
a two-by-two matrix representation for each detector's individual Hilbert spaces. We will follow the same convention as  in \cite{JormaEduPRL},  in which the ground $\ket{g}$ and excited $\ket{e}$ states of the two-level detectors correspond to
\begin{align}
\ket{g}=\left (\!
\begin{array}{c}
1  \\
 0 \\
\end{array}
\!\right ),
\ 
\ket{e}=\left (\!
\begin{array}{c}
0  \\
1 \\
\end{array}
\!\right ),
\ 
\label{eq:mu-matrixrep}
\end{align} 
and the monopole moments
\begin{equation}
m_\nu(\tau) = 
\left (\!
\begin{array}{c c}
0 & e^{-\ii \Omega_\nu \tau} \\
e^{\ii \Omega_\nu \tau} & 0 \\
\end{array}
\!\right )
\, . 
\end{equation}

 In this representation, the most general uncorrelated initial state for the two detectors is
 \begin{align}
\op \rho_{\text{d},0}&=\op\rho\da\otimes\rho\db=\begin{pmatrix}
\alpha_\textsc{a} & \beta_\textsc{a}\\
\beta^*_\textsc{a} &1-\alpha_\textsc{a}
 \end{pmatrix}\otimes\begin{pmatrix}
\alpha_\textsc{b} & \beta_\textsc{b}\\
\beta^*_\textsc{b} &1-\alpha_\textsc{b}
 \end{pmatrix}, \end{align}
where $\alpha_\nu\in\mathbb{R}, \beta_\nu\in\mathbb{C}$  satisfying the conditions that make $\op \rho_{\text{d},0}$ a positive operator. One can now trivially evaluate
\begin{equation}\label{refoo}
\tr[\op m_\textsc{a}(t)\op\rho_{\textsc{a}}]=2\text{Re}\left(\beta\da\,e^{\ii\Omega\da t}\right)
\end{equation}
and also the commutator 
\begin{align}\label{refow}
\big[\op m_\textsc{b}(t'),\op\rho_{\textsc{b}}\big]=\begin{pmatrix}
-2\ii\, \text{Im}\big(\beta\db e^{\ii\Omega\db t'} \big)&e^{-\ii \Omega\db t'} (1-2\alpha\db)\\
-e^{\ii \Omega\db t'} (1-2\alpha\db) & 2\ii\, \text{Im}\big(\beta\db e^{\ii\Omega\db t'} \big)
\end{pmatrix}
\end{align}
Substituting \eqref{refoo} and \eqref{refow} in \eqref{traceda} we get
\begin{align}\label{almostthere}
\nonumber\op\rho_{\textsc{b},\text{signal}}^{(2)}=&\,4 \int_{-\infty}^{\infty}\!\!\!\text{d}t \int_{-\infty}^{\infty}\!\!\!\text{d}t' \, \chi_\textsc{a}(t)\chi_\textsc{b}(t')\text{Re}\left(\beta\da\,e^{\ii\Omega\da t}\right)\\
 &\times\text{Im}\bigg[ W[x_\textsc{a},t,x_\textsc{b},t'] \bigg]\\
\nonumber&\times \begin{pmatrix}
2\, \text{Im}\big(\beta\db e^{\ii\Omega\db t'} \big)&\ii e^{-\ii \Omega\db t'} (1-2\alpha\db)\\
-\ii e^{\ii \Omega\db t'} (1-2\alpha\db) & -2\, \text{Im}\big(\beta\db e^{\ii\Omega\db t'} \big)
\end{pmatrix}
\end{align}
This expression can still be further simplified, since the imaginary part of the Wightman function can be expressed in terms of the expectation value of the field commutator:
\begin{align}\label{commemerges}
\nonumber &\text{Im}\Big(\Tr\big[ \op \phi\bigl({\bm x},t\bigr)\op \phi\bigl({\bm x'},t'\bigr)\op\rho_{\op\phi,0}\big]\Big)\\
\nonumber&=\frac{1}{2\ii}\Big(\Tr\big[ \op \phi\bigl({\bm x},t\bigr)\op \phi\bigl({\bm x'},t'\bigr)\op\rho_{\op\phi,0}\big]-\Tr\big[ \op\rho_{\op\phi,0} \op \phi\bigl({\bm x'},t'\bigr)\op \phi\bigl({\bm x},t\bigr)\big]\Big)\\
\nonumber&=\frac{1}{2\ii}\Tr\big(  \big[\op \phi\bigl({\bm x},t\bigr)\op \phi\bigl({\bm x'},t'\bigr) -\op \phi\bigl({\bm x'},t'\bigr)\op \phi\bigl({\bm x},t\bigr) \big]\op\rho_{\op\phi,0}\big)\\
&=\frac{1}{2\ii}\left\langle\left[\op \phi\bigl({\bm x},t\bigr),\op \phi\bigl({\bm x'},t'\bigr)\right]\right\rangle.
\end{align}
Notice that this quantity is independent of the initial state of the field $\op\rho_{\op\phi,0}$   since the field commutator is a \mbox{c-number}.

Now substituting \eqref{whightman1} into \eqref{almostthere} and using the identity \eqref{commemerges} we can finally write
\begin{align}\label{finalres}
\nonumber\op\rho_{\textsc{b},\text{signal}}^{(2)}&=2 \int_{-\infty}^{\infty}\!\!\!\text{d}t \int_{-\infty}^{\infty}\!\!\!\text{d}t' \, \chi_\textsc{a}(t)\chi_\textsc{b}(t')\text{Re}\left(\beta\da\,e^{\ii\Omega\da t}\right) \mathcal{C}(t,t')\\
&\times\begin{pmatrix}
-2\, \text{Im}\big(\beta\db e^{\ii\Omega\db t'} \big)&-\ii e^{-\ii \Omega\db t'} (1-2\alpha\db)\\
\ii e^{\ii \Omega\db t'} (1-2\alpha\db) & 2\, \text{Im}\big(\beta\db e^{\ii\Omega\db t'} \big)
\end{pmatrix}
\end{align}
where $\mathcal{C}(t,t')$ is the final responsible for the causal behaviour of the leading-order signalling, and it is  simply the integral of the field commutator expectation over the detector's smearing
\begin{align}\label{causalfunctional}
\mathcal{C}(t,t')=&\,\ii\!\!\int\! \text{d}^n\bm x\!\!\int\! \text{d}^n\bm x' F[\bm x\!-\!\bm x_\textsc{a}] F[\bm x'\!-\!\bm x_\textsc{b}]\\
&\nonumber\qquad\qquad\qquad\qquad\qquad\times\left\langle\left[\op \phi\bigl({\bm x},t\bigr),\op \phi\bigl({\bm x'},t'\bigr)\right]\right\rangle.
\end{align}
Notice that \eqref{finalres} is traceless, as it should be the case for all the different order perturbative contributions to the evolved density matrix of detector B. This stems from the fact that all the Dyson series perturbative corrections to the time evolved density matrix  at given order in perturbation theory are traceless (see e.g. \cite{Robort} for a proof). 

The full-density matrix leading-order expressions in Eq. \eqref{finalres} generalize for arbitrary smearings the results obtained in \cite{Robort2} for the leading-order contribution to the transition probability of pointlike detectors. They also generalize the results obtained in \cite{Robort} for pointlike detectors to smeared profiles. We notice already several interesting aspects of \eqref{finalres}: First, all the leading-order contributions to the full density matrix of B coming from the presence of A are proportional to the smeared integral of the expectation of the field commutator, which is a c-number and, therefore, independent of the initial state of the field. The fact that the leading-order signalling contribution to the transition probability of a detector was independent of the background noise  was already noted in \cite{Robort2}. Here we show that this feature is not limited to transition probabilities  and carries over to the full state of the detector, including the detector's quantum coherences, and for arbitrary spatial smearing. Note, however, that this does not necessarily mean that it is equally easy to communicate (at leading order) using two-level detectors independently of the level of `noise' in the field. This is not so because the noise terms in \eqref{eqnoissig} (which give rise to contributions to $\rho_{\textsc{b},T}$ proportional to $\lambda_\textsc{b}^2$) do indeed depend on the background field state. Hence, the signal-to-noise ratio will, in general, depend on the field background state.

Second, since the field commutator always vanishes  for events that are spacelike separated, it is obvious from \eqref{finalres} and \eqref{causalfunctional} that for compactly supported smearing functions, and if no UV cutoffs are introduced in the response of the detector, the time evolution of the full density matrix of detector B is causal, as noted in a slightly different context in \cite{BenicasaBorstenBuckDowker}.  This is also an explicit leading-order extension for compactly supported smeared detectors of the results in all orders of perturbation theory in \cite{mathieuachim1}, which showed that, for pointlike detectors, the microcausality of the scalar theory already guarantees that the  flow of information from detector A to detector B is causal. In other words, here we see explicitly that if the two detectors are compactly smeared and spacelike separated during their entire interaction time, detector B is  not  sensitive  to  the  state  in  which detector A was prepared or even to the very existence of detector A.

Finally, note that \eqref{finalres} under the assumption \eqref{condos} is better behaved in terms of divergences than the noise terms in \eqref{eqnoissig}. This is remarkable because it is well known that the second-order contributions to the time evolved density matrix present UV divergences for three spatial dimensions when the switching is discontinuous \cite{Louko2008}. This divergence is present in the $\lambda_\textsc{b}^2$ noise terms, but it does not appear in the signalling contribution, even in 3+1-dimensional spacetimes and for pointlike detectors plus sudden switching [to see it one can check the expression of the field commutator in 3+1 spacetime dimensions \eqref{3Dcomm} and how it enters in \eqref{finalres}]. We shall discuss this point further in section \ref{seccall}, where we will explicitly consider the pointlike limit of a scenario where both detectors are suddenly switched on and off in 3+1 dimensions.


\section{Effect of non-compact smearing on the causality of detector models}

\subsection{Explicit expressions of $\mathcal{C}(t,t')$ for Gaussian smearing}
We have discussed how the signalling between the two detectors is strictly zero for compactly supported spacelike separated detectors due to the presence of the field commutator in \eqref{finalres}. In this section we will see how one of the most common non-compact smearing profiles (namely Gaussian) affects the causality of the model in the interaction of two particle detectors with the field.

Let us consider the following spatial smearing
\begin{equation}\label{profilee}
F(\bm x)=\frac{1}{(\sigma\sqrt{\pi})^n}e^{-\bm x^2/\sigma^2}
\end{equation}
where $n$ is the number of spatial dimensions. Let us also recall the value of the field commutator in flat spacetime for 1, 2 and 3 spatial dimensions (calculated in full detail in appendix \ref{apA}):
\begin{equation}
\cnm{\op\phi(  x, t)}{\op\phi(   x', t')}=\frac{\ii}{2}\text{sgn}( t'-  t)\Theta\left(| t-  t'|-| x-  x'|\right)
\end{equation}
for 1+1 dimensions,
\begin{align}
&\cnm{\op\phi(\bm x, t)}{\op\phi(\bm  x', t')}=\nonumber\\
&\qquad\frac{\ii}{2\pi}\frac{\text{sgn}( t'-  t)}{\sqrt{( t-  t')^2-|\bm x-\bm  x'|^2}}\Theta\big[( t-  t')^2-|\bm x-\bm  x'|^2\big]
\end{align}
for 2+1 dimensions, and 
\begin{align}\label{callcomm}
&\cnm{\op\phi(\bm x, t)}{\op\phi(\bm  x', t')}=\\
&\quad\frac{\ii}{|\bm x-\bm  x'|}\frac{1}{4\pi}\Big[\delta ( t-  t'+ | \bm x-\bm  x'|) -\delta( t-  t'- | \bm x-\bm  x'|) \Big]\nonumber
\end{align}
for the 3+1-dimensional case (note that, as operators in the Hilbert space of the field, they are all multiples of the identity but we have not made this explicit to relax the already heavy notational load of this paper).

With these ingredients we can evaluate \eqref{causalfunctional} for the different flat spacetime dimensionalities. Let us consider that the two detectors are at rest and their centres of mass are located respectively at $\bm x_\textsc{a}$ and $\bm x_\textsc{b}$. Without loss of generality, we are going to assume that $\bm x_\textsc{a}=0$ and therefore $|\bm x_\textsc{b}|=L$ is the distance between the two detectors' centres of mass. 

\subsubsection{Gaussian smearing in 1+1 dimensions}

For the 1+1-dimensional case, an explicit evaluation of \eqref{causalfunctional} yields
\begin{align}\label{causalfunctional1D}
\mathcal{C}_{1}(t,t')=&\,-\frac{1}{2\pi\sigma^2}\!\!\int_{-\infty}^\infty\!\!\!\!\! \text{d}x\!\!\int_{-\infty}^\infty\!\!\!\!\! \text{d} x' e^{- x^2/\sigma^2} e^{-\bm (x'-L)^2/\sigma^2}\\
&\qquad\qquad\qquad\qquad\qquad\times \Theta\left(| t-  t'|-|x- x'|\right)\nonumber
\end{align}
we can make the following change of variables
\begin{equation}\label{cv}
u=x+x',\qquad v=x-x',
\end{equation}
which makes the Heaviside function easier to handle. We evaluate this expression analytically recalling that, in our setting, detector B will be switched on after detector A [see \eqref{condos} and \eqref{finalres}]. Taking this into account we  obtain
\begin{align}\label{causalfunctional1D2}
\mathcal{C}_{1}(t,t')&=\frac{-1}{4\pi\sigma^2}\!\!\int_{-\infty}^\infty\!\!\!\!\!\! \text{d}u\!\int_{t-t'}^{t'-t}\!\!\!\!\!\!\!\! \text{d} v\, e^{- \frac{(u+v)^2}{(2\sigma)^2}} e^{-\frac{(u-v-2L)^2}{(2\sigma)^2}}\nonumber\\
&\!\!\!\!=\frac{1}{4}\left[\text{Erf}\left(\frac{t-t'+L}{\sqrt{2}\sigma}\right)-\text{Erf}\left(\frac{t'-t+L}{\sqrt{2}\sigma}\right)\right]
\end{align}

\subsubsection{Gaussian smearing in 2+1 dimensions}
For the 2+1-dimensional case
\begin{align}\label{causalfunctional2D}
\mathcal{C}_{2}(t,t')=&\,-\frac{1}{2\pi^3\sigma^4}\!\!\int\!\! \text{d}^2\bm x\!\int\!\! \text{d}^2 \bm x' e^{- \bm x^2/\sigma^2} e^{- (\bm x'-\bm x_\textsc{b})^2/\sigma^2}\nonumber\\
&\qquad\qquad\qquad\times\frac{\Theta\big[( t-  t')^2-|\bm x-\bm  x'|^2\big]}{\sqrt{( t-  t')^2-|\bm x-\bm  x'|^2}}
\end{align}
we can perform a  change of variables similar to \eqref{cv}
\begin{equation}\label{vecchange}
\bm u=\bm x+\bm x',\qquad \bm v=\bm x-\bm x'
\end{equation}
 yielding
\begin{align}\label{causalfunctional2D2}
\mathcal{C}_{2}(t,t')&=\,\frac{-1}{4\pi^3\sigma^4}\!\!\int\!\text{d}^2\bm u\!\!\int \! \text{d}^2\bm v\, e^{- \frac{(\bm u+\bm v)^2}{(2\sigma)^2}} e^{-\frac{(\bm u-\bm v-2\bm x_\textsc{b})^2}{(2\sigma)^2}}\nonumber\\
&\qquad\qquad\qquad\times\frac{\Theta\big[( t-  t')^2-|\bm v|^2\big]}{\sqrt{( t-  t')^2-|\bm v|^2}}
\end{align}
which can be simplified to
\begin{align}\label{causalfunctional2D3}
\mathcal{C}_{2}(t,t')&=-\frac{e^{-L^2/\sigma^2}}{4\pi^3\sigma^4}\!\!\int\! \text{d}^2\bm u\, e^{-\frac{|\bm u|^2-2\bm u\cdot\bm x_\textsc{b}}{2\sigma^2}}\!\!\int \!\! \text{d}^2\bm v\,  e^{-\frac{|\bm v|^2+2\bm v\cdot \bm x_\textsc{b}}{2\sigma^2}}\nonumber\\
&\qquad\qquad\qquad\times\frac{\Theta\big[( t-  t')^2-|\bm v|^2\big]}{\sqrt{( t-  t')^2-|\bm v|^2}}
\end{align}
The integral over $\text{d}^2\bm u$ can  be readily solved analytically
\begin{align}
&\int_{0}^\infty\!\!\! \text{d}|\bm u| |\bm u| e^{-\frac{|\bm u|^2}{2\sigma^2}}\int_0^{2\pi}\!\!\!\text{d}\varphi\,e^{\frac{|\bm u|L \cos\varphi}{\sigma^2}}\nonumber\\
&\quad=2\pi\int_{0}^\infty\!\!\! \text{d}|\bm u| |\bm u| e^{-\frac{|\bm u|^2}{2\sigma^2}}\,I_0\left(\frac{L|\bm u|}{\sigma^2}\right)=2\pi \sigma^2 e^{\frac{L^2}{2\sigma^2}}
\end{align}
where in the intermediate step $I_0(x)$ is the 0-th modified Bessel function of the first kind. Substituting in \eqref{causalfunctional2D3} we get
\begin{align}\label{causalfunctional2D4}
\mathcal{C}_{2}(t,t')&=-\frac{e^{-\frac{L^2}{2\sigma^2}}}{2\pi^2\sigma^2}\!\! \int_0^{t'-t} \!\!\!\!\!\!  \frac{ \text{d}|\bm v|  |\bm v| e^{-\frac{|\bm v|^2}{2\sigma^2}}   }{\sqrt{( t-  t')^2-|\bm v|^2}} \int_0^{2\pi}\!\!\!\text{d}\varphi\,e^{-\frac{|\bm v| L \cos\varphi}{\sigma^2}}\nonumber\\
&= -\frac{e^{-\frac{L^2}{2\sigma^2}}}{\pi\sigma^2}\!\! \int_0^{t'-t} \!\!\!\!\!\!  \frac{ \text{d}|\bm v|  |\bm v| e^{-\frac{|\bm v|^2}{2\sigma^2}}   }{\sqrt{( t-  t')^2-|\bm v|^2}}     I_0\left(\frac{L|\bm v|}{\sigma^2}\right)
\end{align}
which unfortunately,  to the best of the author knowledge, does not admit a closed expression in terms of well-known functions. It is nevertheless regular and easy to evaluate numerically.

\begin{figure*}
\includegraphics[width=0.335\textwidth]{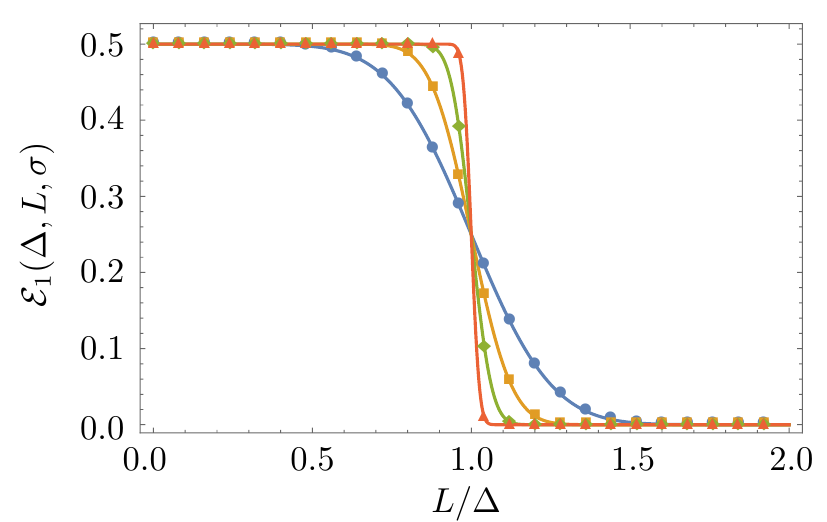}\!\!\!
\includegraphics[width=0.335\textwidth]{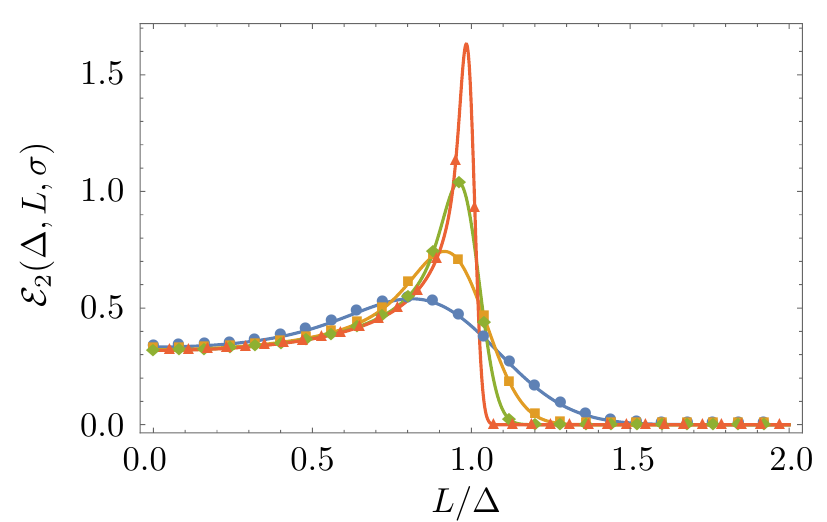}\!\!\!
\includegraphics[width=0.335\textwidth]{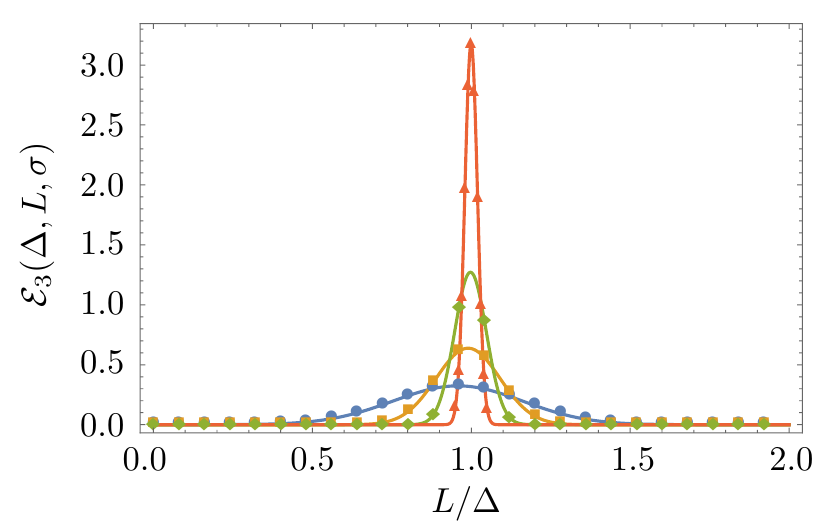}
\caption{Estimator $\mathcal{E}_n(\Delta,L,\sigma)$ of the signalling contribution of the presence of detector A to the state of detector B for (from left to right) $n=1,2$ and 3 dimensions as a function of their spatial separation $L$ (in units of the time separation between their interactions with the field, $\Delta$) and for various characteristic leghtscales $\sigma$ of the Gaussianly smeared detectors. The maximum light-contact between the two detectors occurs when the two detectors are fully null-separated $L=\Delta$. The different lines correspond to values of $\sigma=0.02\Delta$ (red triangles), $\sigma=0.05\Delta$ (green rhombi), $\sigma=0.1\Delta$ (orange squares), $\sigma=0.2\Delta$ (blue circles). We see that in the pointlike limit the estimator vanishes for spacelike separation between detectors. The timelike signalling is non-vanishing for timelike separation in 1+1D and 2+1D showing the violations of the strong Huygens principle. }
\label{fig:three-first}
\end{figure*}

\subsubsection{Gaussian smearing in 3+1 dimensions}

For the 3+1-dimensional case we get that
\begin{align}\label{causalfunctional3D}
\mathcal{C}_{3}(t,t')=&\,-\frac{1}{4\pi^4\sigma^6}\!\!\int\!\! \text{d}^3\bm x\!\int\!\! \text{d}^3 \bm x' e^{- \bm x^2/\sigma^2} e^{- (\bm x'-\bm x_\textsc{b})^2/\sigma^2}\nonumber\\
&\!\!\!\!\!\!\!\!\!\!\!\!\!\!\!\!\!\!\!\!\!\!\!\!\!\times\frac{1}{|\bm x-\bm  x'|}\Big[\delta ( t-  t'+ | \bm x-\bm  x'|) -\delta( t-  t'- | \bm x-\bm  x'|) \Big]
\end{align}
we can again perform the  change of variables \eqref{vecchange}:
\begin{equation}
\bm u=\bm x+\bm x',\qquad \bm v=\bm x-\bm x'
\end{equation}
we also have to disregard the contribution of one of the deltas recalling that, in our setting, detector B will be switched on after detector A (see \eqref{condos} and \eqref{finalres}) so that in the whole integral evaluation of  \eqref{finalres} only $t'>t$ yields non-vanishing contributions:
\begin{align}\label{causalfunctional3D2}
\mathcal{C}_{3}(t,t')&=\,\frac{-1}{8\pi^4\sigma^6}\!\!\int\!\!\text{d}^3\bm u\!\!\int \!\! \text{d}^3\bm v\, e^{-\frac{(\bm u+\bm v)^2}{(2\sigma)^2}} e^{-\frac{(\bm u-\bm v-2\bm x_\textsc{b})^2}{(2\sigma)^2}}\nonumber\\
&\qquad\qquad\qquad\times\frac{1}{|\bm v|}\delta ( t-  t'+ | \bm v|) 
\end{align}
which can be simplified to
\begin{align}\label{causalfunctional3D3}
\mathcal{C}_{3}(t,t')&=-\frac{e^{-L^2/\sigma^2}}{8\pi^4\sigma^6}\!\!\int\!\!\text{d}^3\bm u\, e^{-\frac{|\bm u|^2-2\bm u\cdot\bm x_\textsc{b}}{2\sigma^2}}\!\!\int \!\! \text{d}^3\bm v\,  e^{-\frac{|\bm v|^2+2\bm v\cdot \bm x_\textsc{b}}{2\sigma^2}}\nonumber\\
&\qquad\qquad\qquad\times\frac{1}{|\bm v|}\delta ( t-  t'+ | \bm v|) 
\end{align}
The integral over $\text{d}^3\bm u$ can readily be solved analytically
\begin{align}
&2\pi\int_{0}^\infty\!\!\! \text{d}|\bm u| |\bm u|^2 e^{-\frac{|\bm u|^2}{2\sigma^2}}\int_{-1}^{1}\!\!\!\text{d}(\cos\theta)\,e^{\frac{|\bm u|L \cos\theta}{\sigma^2}}\nonumber\\
&\quad=\frac{4\pi\sigma^2}{L}\int_{0}^\infty\!\!\! \text{d}|\bm u| |\bm u| e^{-\frac{|\bm u|^2}{2\sigma^2}}\sinh\left(\frac{L |\bm u|}{\sigma^2}\right)=2\sqrt{2\pi^3} \sigma^3 e^{\frac{L^2}{2\sigma^2}}
\end{align}
substituting in \eqref{causalfunctional3D3} and performing the integration over $\text{d}^3\bm v$ we get
\begin{align}\label{causalfunctional3D4}
\mathcal{C}_{3}(t,t')&=-\frac{e^{-\frac{L^2}{2\sigma^2}}}{2\sqrt{2 \pi^3}\sigma^3}\!\!\int_{0}^\infty\!\!\!\!\!\text{d}|\bm v| \,|\bm v|  e^{-\frac{|\bm v|^2}{2\sigma^2}}\delta ( t-  t'+ | \bm v|)\nonumber\\
&\times \int_{-1}^{1}\!\!\!\text{d}(\cos\theta)\,e^{\frac{-|\bm v|L \cos\theta}{\sigma^2}}\nonumber \\
&\!\!\!\!\!=\frac{-e^{-\frac{L^2}{2\sigma^2}}}{\sqrt{2 \pi^3}\sigma L}\!\!\int_{0}^\infty\!\!\!\!\!\text{d}|\bm v| \,  e^{-\frac{|\bm v|^2}{2\sigma^2}}\delta ( t-  t'+ | \bm v|)\sinh\left(\frac{L|\bm v|}{\sigma^2}\right) \nonumber\\
&=\frac{-e^{-\frac{L^2}{2\sigma^2}}}{\sqrt{2 \pi^3}\sigma L}  e^{-\frac{(t'-t)^2}{2\sigma^2}}\sinh\left(\frac{L (t'-t)}{\sigma^2}\right) 
\end{align}
which can be further simplified to 
\begin{align}\label{causalfunctional3D5}
\mathcal{C}_{3}(t,t')&=\frac{-1}{2\sqrt{2 \pi^3}\sigma L} \left( e^{-\frac{[L-(t'-t)]^2}{2\sigma^2}}-  e^{-\frac{[L+(t'-t)]^2}{2\sigma^2}}\right)
\end{align}
where the peak at the light cone is explicit.

\subsection{Causality with a Delta switching}

It is possible to do a clean analytic study of the causal response of the Gaussian smeared detector model considering detectors that interact with the field instantaneously. Namely, we consider that detector A perturbs the field only at an instant $t_\textsc{a}$ and detector B switches its interaction at an instant $t_\textsc{b}$. This technically translate into assuming that the switching functions are given by
\begin{equation}
\chi_\textsc{a}(t)=\delta(t-t_\textsc{a}),\qquad \chi_\textsc{b}(t)=\delta(t-t_\textsc{b})
\end{equation}
where, according to the setup that we have been considering $t_\textsc{b}>t_\textsc{a}$. With these switching functions, the leading-order signalling contribution from the presence of Alice to the time-evolved density matrix of Bob is given by
\begin{align}\label{deltafinal}
\nonumber\op\rho_{\textsc{b},\text{signal}}^{(2)}&=2  \text{Re}\left(\beta\da\,e^{\ii\Omega\da t\da}\right) \mathcal{C}(t\da,t\db)\\
&\times\begin{pmatrix}
-2\, \text{Im}\big(\beta\db e^{\ii\Omega\db t\db} \big)&-\ii e^{-\ii \Omega\db t\db} (1-2\alpha\db)\\
\ii e^{\ii \Omega\db t\db} (1-2\alpha\db) & 2\, \text{Im}\big(\beta\db e^{\ii\Omega\db t\db} \big)
\end{pmatrix}
\end{align}
and, from \eqref{deltafinal}, we can see that in this case the signalling contribution will in general be zero if and only if $\mathcal{C}(t\da,t\db)$ is zero. We discussed that if the detectors are compactly supported, this will happen when the detectors become spacelike separated, this is, if we define $\Delta=t\db-t\da$ and recall that we defined $L=|\bm x_\textsc{b}-\bm x_\textsc{a}|$, for pointlike supported detectors we  have that  
\begin{equation}
\Delta<L\Rightarrow \mathcal{C}(t\da,t\db)=0.
\end{equation}
This is not the case for non-compact detectors for which $\mathcal{C}(t\da,t\db)$ never vanishes. To estimate how much casuality is violated in the communication between detectors A and B we build the following estimator of the causal influence of A on B in Eq. \eqref{deltafinal}
\begin{equation}\label{estimfo}
\mathcal{E}_n(\Delta,L,\sigma)=|\mathcal{C}(t\da,t\db)|,
\end{equation}
to which all the matrix elements of detector B in \eqref{deltafinal} are proportional, modulo constants of order 1. From \eqref{causalfunctional1D2}, \eqref{causalfunctional2D4}, \eqref{causalfunctional3D5} we get that
\begin{align}\label{estimator1D}
\mathcal{E}_{1}(\Delta,L,\sigma)=\frac{1}{4}\left[\text{Erf}\left(\frac{L+\Delta}{\sqrt{2}\sigma}\right)-\text{Erf}\left(\frac{L-\Delta}{\sqrt{2}\sigma}\right)\right]
\end{align}
for one spatial dimension,
\begin{align}\label{estimator2D}
\mathcal{E}_{2}(\Delta,L,\sigma)= \frac{e^{-\frac{L^2}{2\sigma^2}}}{\pi\sigma^2}\!\! \int_0^{\Delta} \!  \text{d}y\, \frac{  y\, e^{-\frac{y^2}{2\sigma^2}}   }{\sqrt{\Delta^2-y^2}}     I_0\left(\frac{L \, y}{\sigma^2}\right)
\end{align}
for two spatial dimensions  and
\begin{align}\label{estimator3D}
\mathcal{E}_{3}(\Delta,L,\sigma)= \frac{1}{2\sqrt{2 \pi^3}\sigma L} \left( e^{-\frac{(L-\Delta)^2}{2\sigma^2}}-  e^{-\frac{(L+\Delta )^2}{2\sigma^2}}\right)
\end{align}
for three spatial dimensions.

These magnitudes reflect the strength of the signalling from A in the full state of B after interaction, and thus they should be zero when A and B remain causally disconnected, that is, when $\Delta<L$. In Fig. \ref{fig:three-first} we show, for different values of $\sigma$, how the signalling estimator behaves as a function of the spatial separation of the detectors interacting for a finite amount of time $\Delta$. We see how in the pointlike limit $\sigma\rightarrow0$, the model is fully causal and the signalling estimator vanishes when the detectors are spacelike separated, that is, when $L>\Delta$. We observe how for one and two dimensions there is a leakage of the ability of the detectors to communicate into the timelike area of the light-cone. This is due to the violations of the strong Huygens principle \cite{McLenaghan,czapor}. This phenomenon enables the transmission of information without being supported by an energy flow between sender and receiver, as proved in \cite{Robort2}. Interestingly, although flat 3+1-dimensional spacetimes do not allow for this kind of timelike energyless signalling, almost any other non-flat four-dimensional spacetime does, as for example in expanding universes \cite{Blasco:2015eya}. For our purposes, we also see in Fig. \ref{fig:three-first}  how for finite-sized detectors smeared in a non-compact way, the setup suffers from superluminal signalling. We will quantify this violation of causality and see how fast considering $\sigma$ sufficiently small it is possible to recover approximate causality.

\begin{figure}[h]
\includegraphics[width=0.45\textwidth]{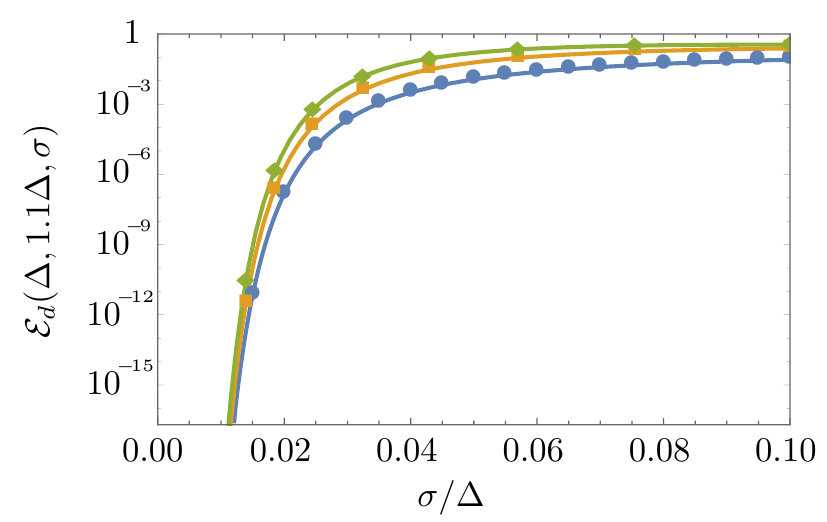}\\[0.3cm]
\includegraphics[width=0.45\textwidth]{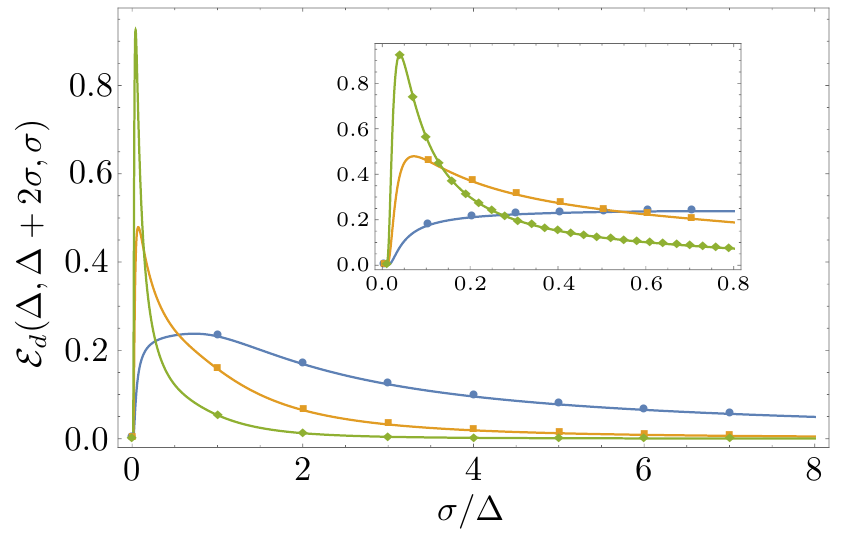}
\caption{{\bf (Top)} Estimator $\mathcal{E}_n(\Delta,L,\sigma)$ of the signalling contribution from detector A to the state of B when the centres of mass of the two detectors are separated by a distance $L=1.1\Delta$ as a function of the smearing lengthscale $\sigma$ (in units of $\Delta$) for $n=$1, 2 and 3 dimensions. We see that the signalling when $\sigma\ll (L-\Delta)$ (which would be zero for compactly supported detectors) decays over-exponentially fast.  {\bf (Bottom)} $\mathcal{E}_n(\Delta,L,\sigma)$  for  $L=\Delta+2\sigma$: when the centres of mass of the two detectors are separated by twice the `size' of the detectors from the spacetime point of maximum light-connection. For compactly supported detectors of size $\sigma$ it would imply spacelike separation and thus $\mathcal{E}_n$ should be zero. We see 1) the over-exponential dampening of the signalling when $\sigma\rightarrow0$, and 2) the signalling from A to B peaks at a dimension-dependent value of $\sigma$ before decaying polynomially on  the lengthscale $\sigma$ (in units of $\Delta$). The different lines correspond to the number of spatial dimensions: $n=1$ (blue circles), $n=2$ (orange squares), $n=3$ (green rhombi). The inset shows a zoomed-in area for $\sigma/\Delta<0.8$.}
\label{figgencaus1}
\end{figure}

This is done in the analysis performed in Fig. \ref{figgencaus1}. In the top panel of Fig. \ref{figgencaus1} we see that for a given detector spatial separation $L=1.1\Delta$ (where $\Delta$ is the spatial separation) the undesired faster-than-light signalling is reduced faster than exponentially as $\sigma\rightarrow0$ regardless of the number of spatial dimensions. Thus, independently of the dimensionality of spacetime, it is possible to recover an approximately causal model for the quantum communication between two detectors when $\sigma\ll L-\Delta$ to arbitrary precision when considering non-compactly smeared detectors.

On a further step, we analyze a crucially different case: in the bottom panel of Fig. \ref{figgencaus1} we consider a scenario where two detectors are separated by a $\sigma$-dependent distance. In particular we consider that $L=\Delta+2\sigma$. That is, the detectors are separated by a distance always equal to twice their characteristic size $\sigma$ from the lightcone. If the detectors were compactly supported with size $\sigma$ this would mean the detectors would be spacelike separated but still close to each other in relation with their own size. The reason for this analysis is to understand if in order for the causality violations to stay in check it is also necessary that the spatial separation is bigger than the detectors' characteristic length, that is, whether we need $L\gg\sigma$ in order to recover approximate causality or not.

We see in the bottom panel of  Fig. \ref{figgencaus1} that, indeed, causality is recovered faster than exponentially as $\sigma$ goes to zero also in this case, as expected from the analysis above. However, there is another aspect of the figure that results interesting and perhaps unexpected at first sight: the causality violations when $L=\Delta+2\sigma$ grow with the size of the detector only up to certain point, which depends on the number of spatial dimensions, before starting a slow polynomial decay as $\sigma$ keeps growing. This perhaps surprising effect stems from the fact that there is a distance decay in the integral of the commutator in the spacelike area of the field in all cases (explicit in 2+1 and 3+1 dimensions but also existent in the 1+1-dimensional case) This decay is not dependent at all on the detector size lengthscale $\sigma$ but it does depend on the dimensionality of spacetime. This means that, for extremely large non-compact supported detectors  whose centres of mass are spacelike separated by only twice their characteristic distance, it is possible to recover some degree of approximate causal behaviour in principle, albeit only very slowly converging to causality as the size of the detectors increases.

\subsection{Dependence on the switching: The regular behaviour of the signalling contributions}\label{seccall}

In the section above we have considered a delta switching corresponding to the instantaneous interaction of the detectors with the field to keep the analysis clean. However, one may legitimately wonder how considering  different compactly supported switching functions (including some finite-time interaction of the detectors with the field) may modify these results about the causality of the Gaussianly smeared detectors.

To convince ourselves that the shape of the switching is not relevant for the results presented in this section, we can analyze the case of an extended switching function in 3+1 dimensions which again yields closed expressions for our signalling estimators. Furthermore, this analysis will allow us to show some interesting aspects about the UV regularity of the signalling contribution \eqref{finalres}.

 For simplicity, let us assume for this subsection that the detectors are gapless. This assumption will remove from the study undesired  oscillations coming from the internal detector dynamics that are irrelevant for the causal behaviour of the detectors. From \eqref{finalres}  we can see that a reasonable signalling estimator for compact switching functions with non-overlapping support can be
\begin{equation}\label{condition}
\mathcal{E_C}= \left|\int_{-\infty}^{\infty}\!\!\!\text{d}t \int_{-\infty}^{\infty}\!\!\!\text{d}t' \, \chi_\textsc{a}(t)\chi_\textsc{b}(t') \mathcal{C}(t,t')\right|.
\end{equation}
Let us consider the following simple sudden switching scenario where the interaction of both detectors has a duration $T$:
\begin{equation}\label{switching}
\chi_\nu(t)=\left\{ \begin{array}{lr}
1 & t\in [t_\nu,t_\nu+T]\\[2mm]
0 & t\notin [t_\nu,t_\nu+T]\\
\end{array}
\right..
\end{equation}
Without loss of generality we consider the switch-on time of detector A $t\da=0$ and we call $\Delta$ the time interval between the switch-off of detector A and the switch-on of detector B. In that case we can express \eqref{condition} for the 3+1-dimensional case, using \eqref{causalfunctional3D5},  as
\begin{equation}\label{condition}
\mathcal{E_C}= \frac{1}{2\sqrt{2 \pi^3}\sigma L}\!\int_{0}^{T}\!\!\!\!\!\text{d}t\!\! \int_{\Delta+T}^{\Delta+2T}\!\!\!\!\!\!\!\!\!\!\!\text{d}t'  \left(\! e^{-\frac{[L-(t'-t)]^2}{2\sigma^2}}\!-\!  e^{-\frac{[L+(t'-t)]^2}{2\sigma^2}}\!\right),
\end{equation}
which admits the following closed form:
\begin{align}\label{exprlarge}
&\mathcal{E_C}(\Delta,L,\sigma,T)=\frac{1}{4 \pi  L}\Big[(L-\Delta ) \text{Erf}\left(\frac{L-\Delta }{\sqrt{2} \sigma }\right)\\
\nonumber&-(\Delta \!+\!L) \text{Erf}\left(\frac{\Delta +L}{\sqrt{2} \sigma }\right)\!+\!2 (\Delta\! -\!L\!+\!T) \text{Erf}\left(\frac{L-\Delta-T}{\sqrt{2} \sigma }\right)\\
\nonumber&+(L-\Delta -2 T) \text{Erf}\left(\frac{ L-\Delta-2 T}{\sqrt{2} \sigma }\right)+2 (\Delta +L+T)\\
\nonumber&\times \text{Erf}\left(\frac{\Delta +L+T}{\sqrt{2} \sigma }\right)-(\Delta +L+2 T) \text{Erf}\left(\frac{\Delta +L+2 T}{\sqrt{2} \sigma }\right)\\
\nonumber&+\sqrt{\frac{2}{\pi }} \sigma  \Big(e^{-\frac{(L-\Delta )^2}{2 \sigma ^2}}\!-\!e^{-\frac{(\Delta +L)^2}{2 \sigma ^2}}\!-\!e^{-\frac{(\Delta -L+T)^2}{2 \sigma ^2}}\!+\!e^{-\frac{(\Delta +L+T)^2}{2 \sigma ^2}}\\
&-\!e^{-\frac{(\Delta -L+T)^2}{2 \sigma ^2}}\!\!+\!e^{-\frac{(\Delta -L+2 T)^2}{2 \sigma ^2}}\!\!+\!e^{-\frac{(\Delta +L+T)^2}{2 \sigma ^2}}\!\!-\!e^{-\frac{(\Delta +L+2 T)^2}{2 \sigma ^2}}\Big)\!\Big].\nonumber
\end{align}

\begin{figure}[h]
\includegraphics[width=0.43\textwidth]{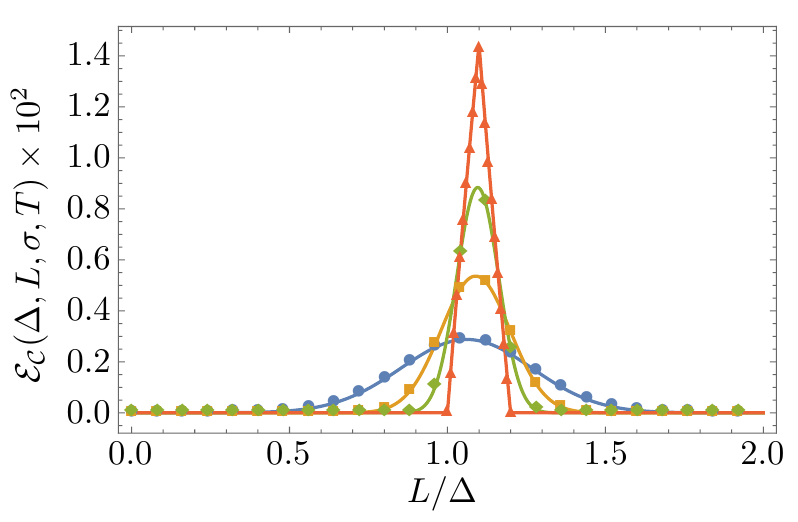}
\caption{ Estimator $\mathcal{E_C}(\Delta,L,\sigma,T)$ of the signalling contribution of the presence of detector A to the state of detector B in a 3+1D spacetime when the detectors are sharply switched on and off after allowing them to interact with the field for a finite time ($T=0.1\Delta$) as a function of their spatial separation $L$ (in units of  $\Delta$) and for various characteristic lenghtscales $\sigma$ of the Gaussianly smeared detectors. The maximum light-contact between the two detectors occurs when $L=\Delta+T$ . The different lines correspond to values of $\sigma=0.02\Delta$ (red triangles), $\sigma=0.05\Delta$ (green rhombi), $\sigma=0.1\Delta$ (orange squares), $\sigma=0.2\Delta$ (blue circles). In the pointlike limit the estimator vanishes for spacelike separation between detectors. The signalling contribution is finite in this limit even with a sharp switching.}
\label{smearedtime}
\end{figure}

In Fig. \ref{smearedtime} we see how the behaviour is qualitatively the same as in the case of a delta switching just accounting for the larger amount of causal contact due to the duration of the interaction $T$. In the figure, the signalling contribution is smaller in magnitude than in the instantaneous switching scenario (which causes a more violent perturbation of the field). Notice an important point: as it can be seen in the plot,  in Fig. \ref{smearedtime}, the pointlike limit $\sigma\rightarrow0$ is no longer a delta distribution, but instead a finite function that has support from the point when the causal contact is initiated at $L=\Delta$, peaks at maximum causal contact $L=\Delta+T$ and decreases down to zero when the causal contact finishes at $L=\Delta+2T$.

We can actually compute the exact limit  $\sigma\rightarrow0$ on \eqref{exprlarge}, which yields
\begin{align}\label{exprlargelim}
&\lim_{\sigma\rightarrow0}\mathcal{E_C}(\Delta,L,\sigma,T)\\
&\qquad=\frac{|L-\Delta|+|L-2T-\Delta |-2|L-T-\Delta|}{4 \pi  L}.\nonumber
\end{align}
The signalling estimator peaks at maximum light-contact $L=\Delta+T$ yielding a finite value, even in the pointlike case, of
\begin{align}\label{exprlargelimpeak}
\mathcal{E_C}^\text{max}=\frac{1}{2 \pi(1+\Delta/T)},
\end{align}
which also does not diverge when the duration of the interaction is considered large $T\rightarrow\infty$.

This is a very remarkable aspect of the leading-order signalling contribution from a detector A to a detector B: It does not diverge when a non-continuous switching function and pointlike detectors are considered. This is in stark contrast to the known divergences  in the noise terms proportional to $\lambda^2\db$ for  3+1-dimensional spacetimes \cite{Louko2008}. What is more, this signalling contribution remains finite and controlled even for long interaction times. This means that, in what concerns the leading-order terms, the signalling contribution can be perturbatively investigated in a consistent way even for very long interaction times. This is however only fully consistent if the larger order corrections to higher power signalling terms were also to remain finite in the limit $T\rightarrow\infty$. Regarding the causal behaviour of the model, we see that the acausal predictions of the non-compact smeared detectors with a finite sharp switching behave in an identical way as the case of the delta switching, reflecting the switching independence of the observed behaviour.

\section{Effect of a UV-cutoff on the causality of communication}

In this section we will analyze the violations of causality in the communication between two particle detectors steaming from the introduction of a hard UV cutoff $\Lambda$   in the interaction between the field and the detector in the  privileged quantization frame $(t,\bm x)$. As discussed in the introduction, it is common to carry this kind of approximations when modelling the light-matter interaction. The rationale of such an approximation is the fact that the coupling between an atom and the electromagnetic field is frequency dependent and will eventually vanish for large enough electromagnetic energies (the atoms become transparent to all frequencies above certain threshold).

This approximation, of course, strongly affects the causal behaviour of the model. In this section we will perform a clean study of the emergent causality violations that appear in the two-detectors communication model when a hard cutoff is introduced. Since we have to have a clean signature of the effect of the UV cutoff and not mix it with the effect of a spatial smearing, we are going to consider delta switching and pointlike detectors for this study in the cases of 1+1, 2+1 and 3+1 dimensions.

The hard UV cutoff is implemented by directly severing the interaction of the field with the detector for field frequencies above a given threshold frequency $\Lambda$. To model this, we need to replace the field operator that couples to the atoms in \eqref{puno} by an operator $\op\phi_\Lambda( x, t)$ which incorporates an UV cutoff $\Lambda$ in its mode expansion. Regarding the detector's causal response, this cutoff involves substituting the field commutators in \eqref{finalres} with UV cutoff versions.

\begin{figure*}
\includegraphics[width=0.343\textwidth]{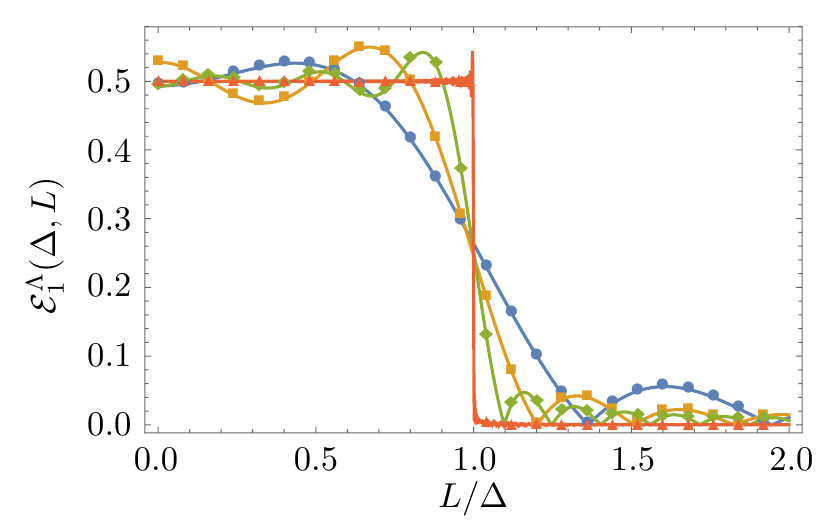}\!\!\!
\includegraphics[width=0.331\textwidth]{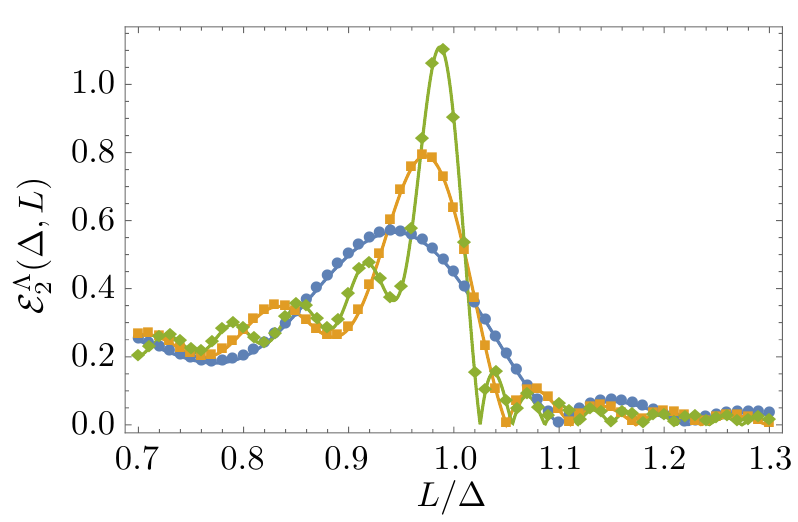}\!\!\!
\includegraphics[width=0.331\textwidth]{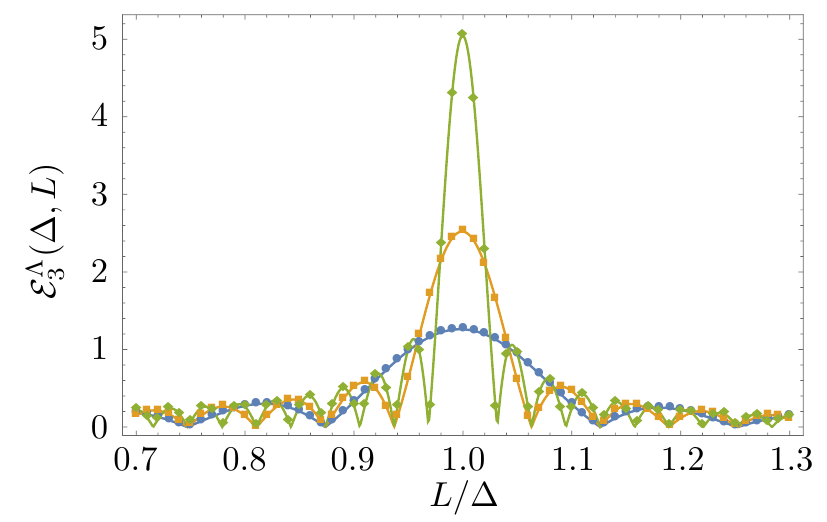}
\caption{Estimator $\mathcal{E}_n^\Lambda(\Delta,L)$ of the signalling contribution of the presence of detector A to the state of detector B for (from left to right) $n=$1,2 and 3 dimensions as a function of their spatial separation $L$ (in units of the time separation between their interactions with the field, $\Delta$) and for various UV cutoffs $\Lambda$. The maximum light-contact between the two detectors occurs when the two detectors are null-separated $L=\Delta$. The different lines of the 1+1D plot correspond to values of $\Lambda=5\Delta^{-1}$ (blue circles), $\Lambda=10\Delta^{-1}$ (orange squares), $\Lambda=20\Delta^{-1}$ (green rhombi) and $\Lambda=10^3\Delta^{-1}$  (orange triangles). For the 2+1D and 3+1D plots they correspond to values of $\Lambda=25\Delta^{-1}$ (blue circles), $\Lambda=50\Delta^{-1}$ (orange squares) and $\Lambda=100\Delta^{-1}$ (green rhombi).  Since the detectors are pointlike and interact with the field instantaneously, the non-vanishing values in the spacelike separation area ($L>\Delta$) are due to causality violations caused by the finite UV-cutoff. Signalling is non-vanishing for timelike separation in 1+1D and 2+1D even in the limit $\Lambda\rightarrow\infty$ due the violations of the strong Huygens principle. }
\label{fig:three-second}
\end{figure*}
\subsection*{1+1 dimensions}

For the case of one spatial dimension we start from \eqref{precutoff1D} and apply the UV cutoff:
\begin{align}
\nonumber&\cnm{\op\phi_\Lambda(  x, t)}{\op\phi_\Lambda(   x', t')}= \\
&=\frac{-\ii}{2\pi}\bigg(\int_{0}^{\Lambda}\!\!\!  \text{d}  k \, \frac{1}{ k} \sin\left[ k ( t-  t'  - x+   x')\right]\\
&\qquad\qquad\qquad+\int_{-\Lambda}^{0}\!\!\!  \text{d}  k \, \frac{1}{- k} \sin\left[ k (- t+ t'  - x+   x')\right]\bigg)\nonumber
\end{align}
The integrals in the expression above are exactly the definition of the sine integral functions, so we can rewrite this UV-cutoff commutator as 
\begin{align}
\nonumber\cnm{\op\phi_\Lambda(  x, t)}{\op\phi_\Lambda(   x', t')}=& \frac{-\ii}{2\pi}\Big(\text{Si}\big[ \Lambda(t-  t'  + x-   x')\big]\\
&-\text{Si}\big[ \Lambda(t-  t'  - x+   x')\big]\Big)
\end{align}
where $\text{Si}(z)$ is the sine integral function.

As mentioned above, we are going to assume that the detectors are pointlike and placed at positions $x\da,x\db$, and that they interact instantaneously with the field at times $t\da$ and $t\db$. Namely,
\begin{equation}
\chi_\textsc{a}(t)=\delta(t-t_\textsc{a}),\  \chi_\textsc{b}(t)=\delta(t-t_\textsc{b}),\ F(x)=\delta(x)
\end{equation}
in Eq. \eqref{finalres}. Without loss of generality, let us assume that $x\da=0,x\db=L$, $t\da=0$ and $t\db=\Delta$. With these choices, we can write the signalling estimator \eqref{estimfo} defined in the previous section as
\begin{align}\label{estimator1DUV}
\mathcal{E}^\Lambda_{1}(\Delta,L)= \frac{1}{2\pi}\left|\text{Si}\big[ \Lambda(L+\Delta)\big]-\text{Si}\big[ \Lambda(L-\Delta)\big]\right|.\end{align}

\subsection*{2+1 dimensions}

We operate under the same hypotheses as in the 1+1-dimensional case (pointlike detectors and delta switchings). Upon the introduction of a UV cutoff in the expression for the field commutator \eqref{precommu2D} and following the same prescriptions as above we can compute the form of the signalling estimator  \eqref{estimfo} for the 2+1-dimensional case
\begin{align}
\mathcal{E}^\Lambda_{2}(\Delta,L)=
&\frac{1}{2\pi}\left|\int_{0}^\Lambda\!\!\! \text{d} y\, J_0\big(y L\big)\sin\big[y  \Delta\big]\right|
\end{align}
which cannot be evaluated i a closed form, but which is again easy to evaluate numerically.

\subsection*{3+1 dimensions}

We repeat the same analysis for the 3+1-dimensional case. From \eqref{precutoff3D}, the cutoff version of the field commutator is
\begin{align}
\cnm{\op\phi_\Lambda(\bm x, t)}{\op\phi_\Lambda(\bm  x', t')}=
&\frac{\ii}{8\pi^2|\bm x-\bm  x'|}\bigg(\int_{-\Lambda}^\Lambda\!\!\!\! \text{d}k \,e^{\ii k ( t-  t'+ | \bm x-\bm  x'|) }\nonumber\\
&-\int_{-\Lambda}^\Lambda\!\! \text{d}k \,e^{\ii k ( t-  t'- | \bm x-\bm  x'|) }\bigg)
\end{align}
which can be readily evaluated to the following closed expression
\begin{align}
&\cnm{\op\phi_\Lambda(\bm x, t)}{\op\phi_\Lambda(\bm  x', t')}=
\frac{\ii}{4\pi^2|\bm x-\bm  x'|}\\
&\!\times\!\bigg(\frac{\sin\big[ \Lambda(t-  t'  \!+\! |\bm x-   \bm x'|)\big]}{t-  t'  + |\bm x-  \bm  x'|}\!+\!\frac{\sin\big[ \Lambda(t'\!-  t  +\! |\bm x-   \bm  x'|)\big]}{t'-  t  + |\bm x-   \bm x'|}\bigg).\nonumber
\end{align}
Again let us call  $L=|\bm x\da-\bm x\db|$, and $\Delta=t\db-t\da$. With these choices, we can write the signalling estimator \eqref{estimfo} for the 3+1-dimensional case as
\begin{align}\label{estimator3DUV}
\mathcal{E}^\Lambda_{3}(\Delta,L)=\frac{1}{4\pi^2L}\bigg|\frac{\sin\big[ \Lambda(L-\Delta)\big]}{L-\Delta}+\frac{\sin\big[ \Lambda(L+\Delta)\big]}{L+\Delta}\bigg|.\end{align}

We show in Fig. \ref{fig:three-second} the behaviour of the estimator $\mathcal{E}_n^\Lambda(\Delta,L)$ as a function of the spatial separation of the detectors $L$ in units of their time separation $\Delta$ for different increasing values of the UV cutoff $\Lambda$ and for 1+1, 2+1 and 3+1 dimensions. We see in which way the fully causal model is approximated as the cutoff increases.

\begin{figure}
\includegraphics[width=0.43\textwidth]{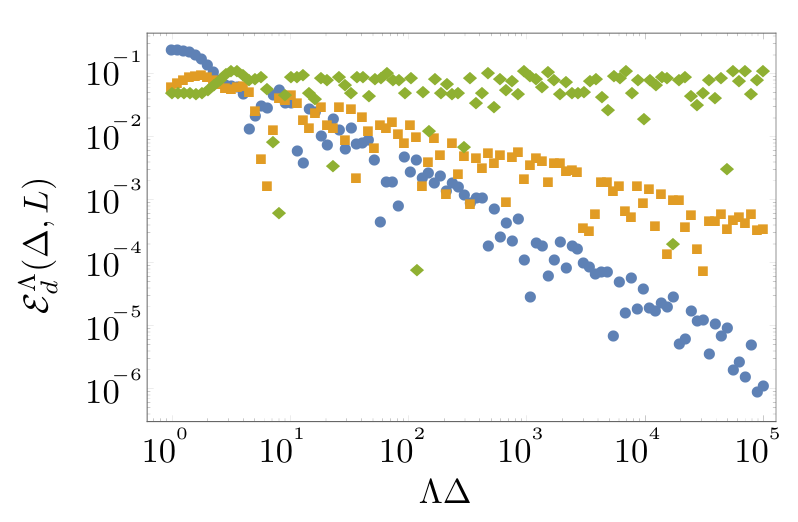}
\caption{ Scatter plot  of the estimator  $\mathcal{E}_n^\Lambda(\Delta,L)$ of the signalling contribution of the presence of detector A to the state of detector B for a fixed spacelike separation of the detectors $L=1.4\Delta$ as a function of the UV cutoff scale $\Lambda$ (in units of $\Delta^{-1}$) for spatial dimensions equal to $n=1$ (blue circles), $n=2$ (orange squares), $n=3$ (green rhombi). In one and two spatial dimensions causality is restored polynomially in $\Lambda$. In three-dimensional spaces increasing the cutoff does not restore causality unless we go to the strict limit $\Lambda\rightarrow\infty$.}
\label{scatterUV}
\end{figure}

One can quickly appreciate a rather concerning aspect of the causal behaviour of this model in the three-dimensional (space) case: we can see from the analytic expression \eqref{estimator3DUV} that increasing the cutoff $\Lambda$ will not fix the acausal signaling for a fixed distance $L$ when considering spacelike separated detectors ($L>\Delta$). Instead of restoring causality, increasing the cutoff will induce faster oscillations, but the magnitude of the acausal influence of A on detector B will remain unaltered and causality will only be recovered in the strict limit $\Lambda\rightarrow\infty$. This can be seen graphically in Fig. \ref{scatterUV} where  we see that increasing the cutoff reduces the acausal influence of A on B at a fixed $L$ polynomially with the UV cutoff for 2+1 and 1+1 dimensions, but it does not reduce it for the 3+1-dimensional case.

In particular, for 2+1 and 1+1 dimensions we see that the acausal influence decreases as $\sim \Lambda^{-\alpha}$, with $\alpha>1$ for 1+1 dimensions and $\alpha<1$ for 2+1 dimensions. These results are consistent with the result for 1+1-dimensional Dirichlet cavities obtained in \cite{Robort} where the causal influence of one detector on another decayed with the square of the number of cavity modes considered in the model.

Although the ill behaviour in 3+1 dimensions may of course be, to some extent, smoothed by considering smoother switchings than a delta-kick, the fact remains that the the restoration of the casual behaviour of the model in 3+1-dimensional spacetimes is much worse than in 1+1 and 2+2 dimensions, and so the use of hard cutoff detector models in three spatial dimensions for detectors switched on for a finite time to scenarios where the causality of the model is relevant is problematic.

We conclude that while it may be argued that approximate causality may be recovered with arbitrary precision polynomially fast in 1+1- and 2+1-dimensional scenarios if we introduce a hard UV cutoff, it is not a reasonable assumption to introduce UV cutoffs in relativistic quantum communication scenarios  in 3+1-dimensional spacetimes.

\section{Causality violations of the rotating-wave approximation}

The rotating-wave approximation (RWA) is yet another very common approximation made in the modelling of quantum optical settings. In fact, it is arguably the most common approximation in quantum optics, and it can be ubiquitously found anywhere from basic textbooks to research works \cite{scullybook}. To better understand this approximation, let us first expand the field in plane-wave modes in the Hamiltonian \eqref{puno}:
\begin{align}\label{fullmodes}
 \op H_I&=\sum_{\nu}\lambda_\nu  \chi_\nu(t)\int \text{d}^n\bm k\,\frac{\tilde F(\bm k)}{\sqrt{|\bm k|}}\\ &\times \Big(\op  a_{\bm k} \op \sigma_\nu^+e^{-\ii[(|\bm k|-\Omega_\nu) t-\bm k\cdot \bm x_\nu]}+\op  a^\dagger_{\bm k} \op \sigma^{-}_\nu e^{\ii[(|\bm k|-\Omega_\nu) t-\bm k\cdot \bm x_\nu]}\nonumber\\
&+\op  a^\dagger_{\bm k} \op \sigma_\nu^+e^{\ii[(|\bm k|+\Omega_\nu) t-\bm k\cdot \bm x_\nu]}+\op  a_{\bm k} \op \sigma^{-}_\nu e^{-\ii[(|\bm k|+\Omega_\nu) t-\bm k\cdot \bm x_\nu]}\Big)\nonumber
\end{align}
where the integral over $\text{d}^n \bm x$ has already been performed and where we have defined the Fourier transform of the real-valued spatial profile as
\begin{equation}
\tilde F(\bm k)=\frac{1}{\sqrt{(2\pi)^n}}\int\text{d}^n\bm x\, F(\bm x) e^{\ii \bm k\cdot\bm x}.
\end{equation}

The RWA consists of removing the terms proportional to $\sigma^+ a^\dagger$ and their Hermitian conjugates in the Hamiltonian \eqref{fullmodes}. More explicitly, the so-called rotating-wave approximation consists of replacing the Hamiltonian \eqref{fullmodes} by
\begin{align}\label{rotatingwave}
\nonumber \op H_I&=\sum_{\nu}\lambda_\nu  \chi_\nu(t)\int \text{d}^n\bm k\,\frac{\tilde F(\bm k)}{\sqrt{|\bm k|}}\\ &\!\times \left(\op  a_{\bm k} \op \sigma_\nu^+e^{-\ii[(|\bm k|-\Omega_\nu) t-\bm k\cdot \bm x_\nu]}+\op  a^\dagger_{\bm k} \op \sigma^{-}_\nu e^{\ii[(|\bm k|-\Omega_\nu) t-\bm k\cdot \bm x_\nu]}\right)\!.
\end{align}
The rationale behind this approximation is that the neglected terms do not have a stationary phase for the detector-field resonance $\bm k=\Omega$, and as such, those terms yield bounded oscillations when integrated in time. Within the approximation, these bounded oscillations can be neglected in the detector-field dynamics when compared to the close-to-resonance rotating-wave terms. Thus, the Hamiltonian \eqref{rotatingwave} is expected to approximate the light-matter interaction for interaction times longer than the Heisenberg time of the detectors $\Omega^{-1}$ \cite{scullybook}.  

Notice that from the form of this Hamiltonian we already see that if the initial state of the field is the vacuum and the detectors are both in the ground state, there is no non-trivial dynamics: the detectors will not be able to communicate or get excited even if the interaction is switched on sharply and for a finite time.

Since the Hamiltonian is no longer linear in the field, the microcausality of the scalar field theory no longer guarantees that the model will behave causally at all. To illustrate and study the causality violations in this approximated model, we are going to consider a particular scenario where we begin with a general state of the two detectors, A and B, but the field is initialized in the vacuum. This is akin to the classical scenario of communication with a pair of antennas in the presence of no background noise,
\begin{equation}
\op\rho_0=\op\rho_{\text{d},0}\otimes\proj{0}{0}.
\end{equation}
To have a clean result and disregard all the other possible sources of acausal behaviour that we have already studied, we are also going to consider pointlike detectors. 

The leading-order signalling will have, again, two contributions corresponding to each term in \eqref{eq:rho2} that is proportional to $\lambda_\textsc{a}\lambda_\textsc{b}$. We will focus on the 3+1-dimensional case and let us evaluate $\tr_\phi\big(\op U^{(1)}\op\rho_0\op {U^{(1)}}^\dagger\big)$:
\begin{align}
&\tr_\phi\big(\op U^{(1)}\op\rho_0\op {U^{(1)}}^\dagger\big)=\frac{1}{16\pi^3}\sum_{\nu,\eta}\lambda_\nu\lambda_\eta\,  \op\sigma_\nu^-\op\rho_{\text{d},0}\op\sigma_\eta^+\\
&\nonumber\times\!\! \int_{-\infty}^\infty\!\!\!\!\!\text{d}t\int_{-\infty}^\infty\!\!\!\!\!\text{d}t'\chi_\nu(t)\chi_\eta(t')\!\int\!\frac{\text{d}^3\bm k}{|\bm k|} e^{\ii[(|\bm k|-\Omega) (t-t')-\bm k\cdot \bm (\bm x_\nu-\bm x_\eta)]}  
\end{align}
where we have used that $\tr_\phi\big(a^\dagger_{\bm k}\proj{0}{0}a_{\bm k'})=\delta^{(3)}(\bm k-\bm k')$ and the fact that the detectors are considered pointlike, that is $\tilde F(\bm k)=1$.

We can quickly evaluate the integral over $\text{d}^3 \bm k$ that we will denote as 
\begin{align}\label{rotwaveca}
&\nonumber\mathcal{C}(t,t',\bm x_\nu,\bm x_\eta)=\int\!\frac{\text{d}^3\bm k}{|\bm k|} e^{\ii[(|\bm k|-\Omega) (t-t')-\bm k\cdot \bm (\bm x_\nu-\bm x_\eta)]}\\
\nonumber&=\frac{4\pi\, e^{-\ii\Omega(t-t')}}{|\bm x_\nu-\bm x_\eta|} \int_0^\infty\!\!\!\text{d}|\bm k|\, e^{\ii |\bm k|(t-t')}\sin\big(|\bm k| |\bm x_\nu-\bm x_\eta|\big)\\
\nonumber&=\frac{4\pi\, e^{-\ii\Omega(t-t')}}{|\bm x_\nu-\bm x_\eta|} \bigg(\frac{|\bm x_\nu-\bm x_\eta|}{|\bm x_\nu-\bm x_\eta|^2-(t-t')^2}+\frac{\ii \pi}{2}\\
&\times\Big[\delta(t-t'-|\bm x_\nu-\bm x_\eta|)-\delta(t-t'-|\bm x_\nu+\bm x_\eta|)\Big]\bigg).
\end{align}
Here we see explicitly how for the signalling terms (proportional to $\lambda_\textsc{a}\lambda_\textsc{b}$), the rotating-wave approximation breaks causality: if we compare this expression with the commutator \eqref{callcomm}, we see that the result under the rotating-wave approximation is essentially the same in terms of causal behaviour than the full model without approximation, except for the first summand in \eqref{rotwaveca}, which is explicitly non-vanishing outside the light-cone and thus introduces causality violations.

Let us compute the signalling contribution from A to the density matrix of B in the terms $\tr_\phi\big(\op U^{(1)}\op\rho_0\op {U^{(1)\dagger}}\big)$ of \eqref{eq:rho2} under the RWA by tracing out detector A in the terms proportional to $\lambda_\textsc{a}\lambda_\textsc{b}$ of \eqref{rotwaveca}:
 \begin{align}
\nonumber&\tr_{\phi,A}\big(\op U^{(1)}\op\rho_0\op {U^{(1)}}^\dagger\big)=\frac{\lambda\da\lambda\db}{16\pi^3}\bigg(  \tr(\op\sigma\da^-\op\rho_{\textsc{a},0})\op\rho_{\textsc{b},0}\op\sigma\db^+\\
&\qquad\qquad\nonumber\times\!\! \int_{-\infty}^\infty\!\!\!\!\!\text{d}t\int_{-\infty}^\infty\!\!\!\!\!\text{d}t'\chi_\textsc{a}(t)\chi_\textsc{b}(t')\,\mathcal{C}(t,t',\bm x\da,\bm x\db)\\
&\qquad\qquad+\tr(\op\rho_{\textsc{a},0}\op\sigma\da^+)\op\sigma\db^-\op\rho_{\textsc{b},0}\\
&\qquad\qquad\nonumber\times\!\! \int_{-\infty}^\infty\!\!\!\!\!\text{d}t\int_{-\infty}^\infty\!\!\!\!\!\text{d}t'\chi_\textsc{b}(t)\chi_\textsc{a}(t')\,  \mathcal{C}(t,t',\bm x\db,\bm x\da) \bigg) . 
\end{align}

It is easy to see that the signalling contribution to the density matrix of B coming from $\op U^{(2)}\op\rho_0+\op\rho_0\op U^{(2)\dagger}$ yields identically behaved terms in terms of the scaling of causality violation as those coming from $\op U^{(1)}\op\rho_0\op {U^{(1)\dagger}}$  under the hypothesis \eqref{condos}.

\begin{figure}
\includegraphics[width=0.45\textwidth]{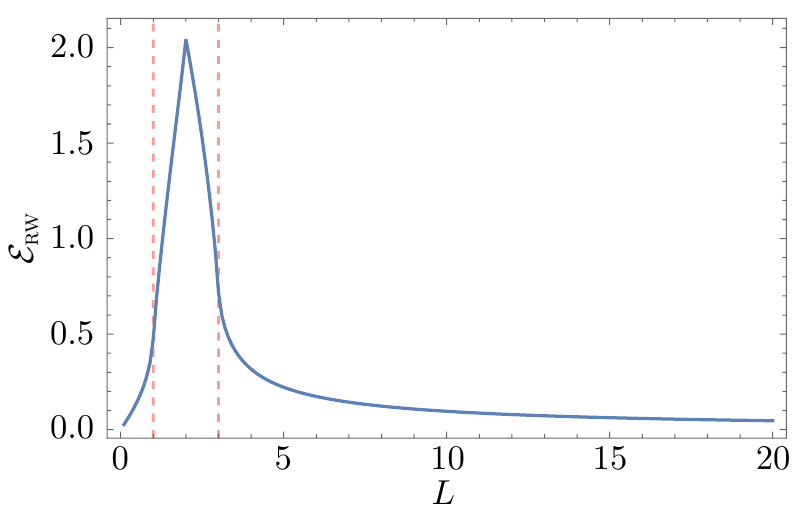}\\[0.3cm]
\includegraphics[width=0.46\textwidth]{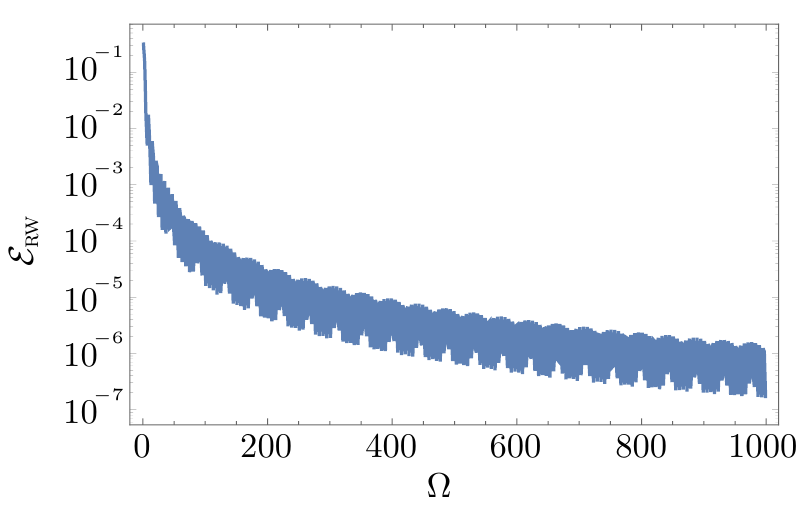}
\caption{{\bf (Top)} Estimator $\mathcal{E}_{RW}$ of the signalling contribution from detector A to the state of B under the RWA for two pointlike detectors  separated by a distance $L$ (in units of $\Omega^{-1}$). The duration of the interaction for both detectors is $T=\Omega^{-1}$ and the interval between A's switching off and B's swithcing on is $\Delta=\Omega^{-1}$. The values of $L$ where the detectors' interactions have some light contact are enclosed by two vertical red dashed lines. The signalling estimator peaks at maximum causal contact. We see that the signalling at strictly spacelike separation ($L>\Delta+2T$) is non-zero due to the causality violation introduced by the RWA, decreasing slowly with the spatial separation.  {\bf (Bottom)}  $\mathcal{E}_{RW}$  for fixed $L=3T+\Delta$ (detectors fully spacelike separated) as a function of $\Omega$ in units of $T^{-1}$: We see that while it is true that causality gets restored when $T\ll\Omega^{-1}$, the causality violation is only polynomially recovered when $\Omega$ grows. }
\label{figgencausRWA}
\end{figure}

Therefore, to estimate the causality violations in the two-detector communication scenario coming from the rotating-wave approximation, we need to assess the magnitude of the following signalling estimator which will be zero if there is no influence of A on B:
\begin{equation}\label{gge}
\mathcal{E}_{\textsc{rw}}=\left|\int_{-\infty}^\infty\!\!\!\!\!\text{d}t\int_{-\infty}^\infty\!\!\!\!\!\text{d}t'\chi_\textsc{a}(t)\chi_\textsc{b}(t') \frac{e^{-\ii\Omega(t-t')}}{L^2-(t-t')^2}\right|
\end{equation}
This is, the terms of the causal influence of A on B (Eq. \eqref{rotwaveca}) that are non-zero outside of the lightcone. Because it is argued that the rotating-wave approximation only becomes accurate for long durations of the interaction, we are going to consider a finite-duration interaction. Namely, we will use the switching function \eqref{switching}. In other words, the interactions of A and B have a duration $T$, A switches on at $t=0$ and there is a delay $\Delta$ between the switching off of A and the switching on of B. This leaves \eqref{gge} as follows
\begin{equation}\label{gge2}
\mathcal{E}_{\textsc{rw}}=\left|\int_{0}^T\!\!\text{d}t\!\int_{T+\Delta}^{2T+\Delta}\!\!\!\!\!\!\!\text{d}t'  \frac{e^{-\ii\Omega(t-t')}}{L^2-(t-t')^2}\right|
\end{equation}
this integral can be evaluated analytically, for which we first perform the change of variables
\begin{equation}\label{cv2}
u=t+t',\qquad v=t-t',
\end{equation}
 which maps the integral to
\begin{align}\label{gge3}
\nonumber\mathcal{E}_{\textsc{rw}}=&\frac{1}{2}\bigg|\int_{T+\Delta}^{2T+\Delta}\!\!\!\!\!\text{d}u\!\int_{-u}^{u-2T-2\Delta}\!\!\!\!\!\!\!\text{d}v\, e^{-\ii\Omega v} \bigg(\frac{1}{L^2-v^2}\bigg)\\
&+\int_{2T+\Delta}^{3T+\Delta}\!\!\!\!\!\text{d}u\!\int_{u-4T-2\Delta}^{-u+2T\Delta}\!\!\!\!\!\!\!\text{d}v\, e^{-\ii\Omega v} \bigg(\frac{1}{L^2-v^2}\bigg)\bigg|
\end{align}
The closed expression is lengthy (a long sum of sine integral and cosine integral functions) but easily evaluable.

The two pointlike detectors are completely spacelike separated for values of $L$ satisfying $L>2T+\Delta$. To see how the violation of causality of the rotating-wave approximation behaves with the increase in the duration of the interaction, we are going to fix $L=3T+\Delta$ so that the detectors are always fully spacelike separated during the interaction.

We see in Fig. \ref{figgencausRWA} that when $T\sim\Omega^{-1}$ (the interaction times where the RWA is usually considered to start  being unacceptable), the causality violation decreases very slowly with the spatial separation betweeen the detectors. The fact that the RWA model allows for long-range faster-than-light signalling renders the model in this regime completely unsafe for any kind of relativistic considerations.

We see also that as we increase the product $\Omega T$ the RWA violation of causality becomes smaller as expected. However, importantly, as $\Omega T$ increases, the speed at which causality is restored in the model becomes increasingly slower: while it is possible to improve the causality of the RWA-approximated model if the interaction time is much larger than $\Omega^{-1}$, the speed at which causality is recovered drops fast as the duration of the interaction increases. While it seems possible to work under the RWA and have some approximate degree of causal behaviour for very long interaction times, the spacelike signalling that the RWA enables is resilient, long-ranged (in terms of the spatial separation of the detectors) and remains even for relatively long timescales. This means that one should be extremely careful when using this approximation in settings where faster-than-light signalling cannot be tolerated. For instance to make predictions about experiments where the outcome relies on the fact that the detectors remain spacelike separated.

\section{Conclusions}

We have analyzed, in an operational way, the causal behaviour of three of the common approximations and generalizations of the particle detector models used in quantum field theory to probe quantum fields (such as the Unruh-DeWitt detector) and of the closely related models of light-matter interaction (such as the Jaynes-Cummings model). Namely, we have studied  1) the introduction of a non-compact spatial smearing of the detector, 2) the introduction of a UV cutoff in the detector-field interaction and 3) the use of the rotating-wave approximation (RWA).

Concretely, we have studied how the three modifications of the model enable unacceptable faster-than-light signalling between two particle detectors interacting  with a quantum field at different times in flat spacetimes of one, two and three spatial dimensions. While all these three modifications of the standard UDW model introduce some degree of pathological faster-than-light signalling in quantum communication scenarios, we have seen that some are more harmful than others. 

First, we have analyzed the causality violations that appear when particle detectors are spatially smeared with a smearing function that is not compactly supported. This is of interest because this kind of approximation is very often considered in the literature, both as a means to regularize divergences or as a refinement of the light-matter interaction models  to include the finite size of the (non-compactly supported) atomic wavefunction. We have seen that the causality violations introduced by a Gaussianly supported spatial profile can be made arbitrarily small over-exponentially fast as we reduce the size of the detectors in units of the time separation between them, regardless of the number of spacetime dimensions.

Second, we have studied the effect of the introduction of a UV cutoff on the signalling between two spacelike separated detectors. This kind of approximation is made in the context of the light-matter interaction and it is often justified by the fact that atomic probes do not couple with the same strength to all the frequencies of the electromagnetic field, being effectively transparent to frequencies much above the atomic frequency resonances. We have seen that the effect of this kind of cutoff is critically dependent on the dimension of spacetime, and so are the limitations on the model that relativistic causality imposes. Namely, while the violation of causality can be made arbitrarily small for 1+1 and 2+1 dimensions polynomially fast on the frequency of the cutoff, the case of 3+1 dimensions is different: for very fast switchings it is not possible to make the faster-than-light signalling arbitrarily small for any finite value of the UV cutoff. This renders this kind of approximation in 3+1 dimensions unsuitable to describe scenarios in relativistic quantum communication or any other scenarios where relativistic considerations are important. 

Finally, we have assessed the violations of causality that emerge from the application of the rotating-wave approximation (commonly used in the light-matter interaction models such as the Jaynes-Cummings model). We have seen that the RWA allows for long-range faster-than-light signalling for interaction times of the order of the detector's Heisenberg time $\Omega^{-1}$. The violations of causality do get reduced as the interaction time becomes larger than the detector's Heisenberg time, but their reduction is slow and that makes the RWA perilous to use  and arguably an approximation to be avoided ---even for long interaction times--- when describing setups in relativistic quantum communication and any other circumstance where the causality of the model is crucial.

Additionally, we have discussed some remarkable aspects of the quantum communication between two particle detectors through a quantum  field. First, the leading-order signalling contributions of one particle detector to the state of another detector (that is, the influence that a detector A switched on at early times has over a detector B switched in A's future) is completely independent of the background initial state of the field regardless of the initial state of detectors A and B. This is commensurate with what was seen for the transition rate of detector B in \cite{Robort2}. Nevertheless, we have discussed that signalling from A to B is still affected by the background state through the local noise terms. Second, we have seen that the influence of detector A on the state of detector B (referred to as the {\it signalling contributions} throughout this work) does not fundamentally present the kind of UV divergences described in \cite{Louko2008} for pointlike detectors and discontinuous switching functions. 

\section{Acknowledgements}

The author has a very special thanks for Achim Kempf for his invaluable continuous support and encouragement.  E. M.-M. is supported by the National Sciences and Engineering Research Council of Canada through the Discovery program.

\appendix

\section{Field commutators in different spacetimes}\label{apA}

In this appendix we obtain in full detail the scalar field commutator in flat spacetime for 1+1, 2+1 and 3+1 spacetime dimensions. Although very basic, these calculations are going to be useful to show how to obtain the field commutator when an UV cutoff is introduced.

Notice that throughout this appendix the results of these commutator evaluations have to be regarded as c-numbers (multiples of the identity in the field's Hilbert space).

\subsection{Free scalar field commutator in 1+1-dimensional Minkowski spacetime}

The easiest way to obtain this commutator is to consider a plane-wave mode expansion of the field:
\begin{align}\op\phi(  x,t)=\int_{-\infty}^{\infty}\!\!\!  \text{d}  k  \frac{1}{\sqrt{4\pi |k|}}\left(\hat a_{  k} e^{-\ii(|  k| t-  k     x)}+\hat a^\dagger_{  k} e^{\ii(|  k| t-  k     x)}\right).\end{align}
Using this we get
\begin{align}
\nonumber&\cnm{\op\phi(  x, t)}{\op\phi(   x', t')}=\\ \nonumber&\frac{1}{4\pi}\int_{-\infty}^{\infty}\!\!\!\!\!\!  \text{d}  k  \int_{-\infty}^{\infty}\!\!\! \!\!\! \text{d}  k ' \frac{1}{\sqrt{|  k||  k'|}}\Big(\cnm{\hat a_{  k}}{\hat a_{  k'}^\dagger} e^{-\ii[(|  k|  t-|  k'|  t')-  k      x-  k'     x')]}\\&+\cnm{\hat a^\dagger_{  k}}{\hat a_{  k'}} e^{\ii[(|  k|  t-|  k'|  t')-  k      x-  k'     x')]}\Big)
\end{align}
where we have used that $\cnm{a_{  k}}{a_{  k'}}=\cnm{a^\dagger_{  k}}{a^\dagger_{  k'}}=0$. Now since 
\begin{equation}
\cnm{a_{  k}}{a^\dagger_{  k'}}=\delta(  k -   k'),\ \cnm{\hat a^\dagger_{  k}}{\hat a_{  k'}}=-\cnm{a^\dagger_{  k'}}{a_{  k}}=-\delta(  k' -   k)
\end{equation}
we have that
\begin{align}\label{precutoff1D}
\nonumber&\cnm{\op\phi(  x, t)}{\op\phi(   x', t')}= \frac{1}{4\pi}\int_{-\infty}^{\infty}\!\!\!  \text{d}  k \, \frac{1}{|  k|}\bigg( e^{-\ii[(|  k| ( t-  t')-  k  (   x-   x')]}\\
&\qquad\qquad\qquad\qquad\qquad\qquad\qquad- e^{\ii[|  k| ( t-  t')-  k  (   x-   x')]}\bigg)\nonumber\\
&=\frac{-\ii}{2\pi}\bigg(\int_{0}^{\infty}\!\!\!  \text{d}  k \, \frac{1}{ k} \sin\left[ k ( t-  t'  - x+   x')\right]\nonumber\\
&\qquad\qquad\qquad+\int_{-\infty}^{0}\!\!\!  \text{d}  k \, \frac{1}{- k} \sin\left[ k (- t+ t'  - x+   x')\right]\bigg)\nonumber\\
\end{align}
Now since $\displaystyle{\int_{0}^\infty \text{d}x\frac{1}{x}\sin(a x)=\frac{\pi}{2}\text{sgn}(a)}$,
\begin{align}
\cnm{\op\phi(  x, t)}{\op\phi(   x', t')}&=\frac{-\ii}{4}\Big(\text{sgn}\big[ t-  t'  - x+   x'\big]\nonumber\\
&\qquad\quad-\text{sgn}\big[ t'-  t  - x+   x'\big]\Big)
\end{align}
which can be simplified to
\begin{equation}
\cnm{\op\phi(  x, t)}{\op\phi(   x', t')}=\frac{\ii}{2}\text{sgn}( t'-  t)\Theta\left(| t-  t'|-| x-  x'|\right)
\end{equation}

\subsection{Free scalar field commutator in 2+1-dimensional Minkowski spacetime}

Again we make a plane-wave mode expansion of the field, as in the 1+1-dimensional case:
\begin{equation}
\op\phi(\bm x,t)=\int  \frac{\text{d}^2\bm k}{\sqrt{(2\pi)^2 2|\bm k|}}\left(\hat a_{\bm k} e^{-\ii(|\bm k| t-\bm k \cdot \bm x)}+\hat a^\dagger_{\bm k} e^{\ii(|\bm k| t-\bm k \cdot \bm x)}\right).
\end{equation}
We can now use this expansion to obtain that
\begin{align}
\nonumber&\cnm{\op\phi(\bm x, t)}{\op\phi(\bm  x', t')}=\\ &\frac{1}{8\pi^2}\!\int \!\text{d}^2\bm k \!\int\! \text{d}^2\bm k' \frac{1}{\sqrt{|\bm k||\bm k'|}}\Big(\cnm{\hat a^\dagger_{\bm k}}{\hat a_{\bm k'}} e^{\ii[(|\bm k|  t-|\bm k'|  t')-\bm k \cdot  \bm x-\bm k'\cdot \bm  x')]}\nonumber\\
&\qquad\qquad\qquad\qquad\quad+\cnm{\hat a_{\bm k}}{\hat a_{\bm k'}^\dagger} e^{-\ii[(|\bm k|  t-|\bm k'|  t')-\bm k \cdot  \bm x-\bm k'\cdot \bm  x')]}\Big)\nonumber
\end{align}
where we have used that $\cnm{a_{\bm k}}{a_{\bm k'}}=\cnm{a^\dagger_{\bm k}}{a^\dagger_{\bm k'}}=0$. Now since $\cnm{a_{\bm k}}{a^\dagger_{\bm k'}}=\delta^{(2)}(\bm k - \bm k')$,
\begin{align}\label{precommu2D}
\nonumber&\cnm{\op\phi(\bm x, t)}{\op\phi(\bm  x', t')}= \frac{1}{8\pi^2}\int \frac{ \text{d}^2\bm k}{|\bm k|}\Big( e^{-\ii[(|\bm k| ( t-  t')-\bm k \cdot( \bm x-\bm  x')]}\\
&\qquad\qquad\qquad\qquad\qquad\qquad- e^{\ii[(|\bm k| ( t-  t')-\bm k \cdot( \bm x-\bm  x')]}\Big)\nonumber\\
&=\frac{1}{8\pi^2}\int_{0}^\infty\!\!\! \text{d}|\bm k|\, \int_{0}^{2\pi}\!\!\!\!\! \text{d}\varphi \Big( e^{-\ii[(|\bm k| ( t-  t')-|\bm k| |\bm x-\bm  x'|\cos\varphi]}\\
&\qquad\qquad- e^{\ii[(|\bm k| ( t-  t')-|\bm k| |\bm x-\bm  x'|\cos\varphi]}\Big)\nonumber \\
&=\frac{1}{2\ii\pi}\int_{0}^\infty\!\!\! \text{d}|\bm k|\, J_0\big(|\bm k| |\bm x-\bm  x'|\big)\sin\big[|\bm k| ( t-  t')\big]
\end{align}
which can be evaluated in closed form as
\begin{align}
&\cnm{\op\phi(\bm x, t)}{\op\phi(\bm  x', t')}=\nonumber\\
&\qquad\frac{\ii}{2\pi}\frac{\text{sgn}( t'-  t)}{\sqrt{( t-  t')^2-|\bm x-\bm  x'|^2}}\Theta\big[( t-  t')^2-|\bm x-\bm  x'|^2\big]
\end{align}

\subsection{Free scalar field commutator in 3+1-dimensional Minkowski spacetime}

Once again, we expand the field in terms of plane-wave modes as we did in the previous cases:
\begin{equation}\hat \phi(\bm x,t)=\int \!\!\frac{\text{d}^3\bm k}{\sqrt{(2\pi)^3 2|\bm k|}}\left(\hat a_{\bm k} e^{-\ii(|\bm k| t-\bm k \cdot \bm x)}+\hat a^\dagger_{\bm k} e^{\ii(|\bm k| t-\bm k \cdot \bm x)}\right).\end{equation}
With this expansion at hand we obtain that
\begin{align}
\nonumber&\cnm{\op\phi(\bm x, t)}{\op\phi(\bm  x', t')}=\\ &\frac{1}{16\pi^3}\!\!\int\!\! \text{d}^3\bm k\! \int\!\! \text{d}^3\bm k' \frac{1}{\sqrt{|\bm k||\bm k'|}}\!\Big(\cnm{\hat a_{\bm k}}{\hat a_{\bm k'}^\dagger} e^{-\ii[(|\bm k|  t-|\bm k'|  t')-\bm k \cdot  \bm x-\bm k'\cdot \bm  x')]}\nonumber\\
&+\cnm{\hat a^\dagger_{\bm k}}{\hat a_{\bm k'}} e^{\ii[(|\bm k|  t-|\bm k'|  t')-\bm k \cdot  \bm x-\bm k'\cdot \bm  x')]}\Big)\nonumber
\end{align}
where we have used that $\cnm{a_{\bm k}}{a_{\bm k'}}=\cnm{a_{\bm k}}{a^\dagger_{\bm k'}}=0$. Now since $\cnm{a_{\bm k}}{a^\dagger_{\bm k'}}=\delta^{(3)}(\bm k - \bm k')$, 
\begin{align}
\nonumber&\cnm{\op\phi(\bm x, t)}{\op\phi(\bm  x', t')}= \frac{1}{16\pi^3}\int \frac{ \text{d}^3\bm k}{|\bm k|}\Big( e^{-\ii[|\bm k| ( t-  t')-\bm k \cdot( \bm x-\bm  x')]}\\
&\qquad\qquad\qquad\qquad\qquad\qquad- e^{\ii[|\bm k| ( t-  t')-\bm k \cdot( \bm x-\bm  x')]}\Big)\nonumber \\
&=\frac{-\ii}{8\pi^3}\int \text{d}^3\bm k\, \frac{1}{|\bm k|}\sin\big[|\bm k| ( t-  t')-\bm k \cdot( \bm x-\bm  x')\big]\nonumber\\
&=\frac{\ii}{4\pi^2}\!\int\!\! \text{d}|\bm k| |\bm k|\!\!\int_{1}^{-1}\!\!\!\!\!\!\text{d}(\cos\theta)\sin\big[|\bm k| [( t-  t')\!-\! | \bm x-\bm  x'|\cos\theta]\big]\nonumber\\
&=\frac{\ii}{4\pi^2}\int \text{d}|\bm k|\, \frac{1}{|\bm x-\bm  x'|}\Big(\cos\left[|\bm k| ( t-  t'+| \bm x-\bm  x'|)\right]\nonumber\\
&\qquad\qquad\qquad\qquad\qquad-\cos\left[|\bm k| ( t-  t'-| \bm x-\bm  x'|)\right]\Big)\nonumber
\end{align}
\begin{align}\label{precutoff3D}
&=\frac{\ii}{|\bm x-\bm  x'|}\frac{1}{8\pi^2}\bigg(\int_{0}^\infty \text{d}|\bm k|\,e^{\ii|\bm k| ( t-  t'+ | \bm x-\bm  x'|) }\nonumber\\
&+\int_{0}^\infty \text{d}|\bm k|\,e^{-\ii|\bm k| ( t-  t'+ | \bm x-\bm  x'|) }-\int_{0}^\infty \text{d}|\bm k|\,e^{\ii|\bm k| ( t-  t'- | \bm x-\bm  x'|) }\nonumber\\
&\qquad\qquad\qquad\qquad\qquad-\int_{0}^\infty \text{d}|\bm k|\,e^{-\ii|\bm k| ( t-  t'- | \bm x-\bm  x'|) }\bigg)\nonumber\\
&=\frac{\ii}{|\bm x-\bm  x'|}\frac{1}{8\pi^2}\bigg(\int_{-\infty}^\infty\!\! \text{d}k \,e^{\ii k ( t-  t'+ | \bm x-\bm  x'|) }\nonumber\\
&\qquad\qquad\qquad\qquad-\int_{-\infty}^\infty\!\! \text{d}k \,e^{\ii k ( t-  t'- | \bm x-\bm  x'|) }\bigg)
\end{align}
which finally yields the following closed expression with support only on the null boundary of the lightcone, (in contrast to the timelike leakage of  1+1D and 2+1D)
\begin{align}\label{3Dcomm}
&\cnm{\op\phi(\bm x, t)}{\op\phi(\bm  x', t')}=\\
&\quad\frac{\ii}{|\bm x-\bm  x'|}\frac{1}{4\pi}\Big[\delta ( t-  t'+ | \bm x-\bm  x'|) -\delta( t-  t'- | \bm x-\bm  x'|) \Big].\nonumber
\end{align}

\bibliography{references}

\begin{thebibliography}{53}%
\makeatletter
\providecommand \@ifxundefined [1]{%
 \@ifx{#1\undefined}
}%
\providecommand \@ifnum [1]{%
 \ifnum #1\expandafter \@firstoftwo
 \else \expandafter \@secondoftwo
 \fi
}%
\providecommand \@ifx [1]{%
 \ifx #1\expandafter \@firstoftwo
 \else \expandafter \@secondoftwo
 \fi
}%
\providecommand \natexlab [1]{#1}%
\providecommand \enquote  [1]{``#1''}%
\providecommand \bibnamefont  [1]{#1}%
\providecommand \bibfnamefont [1]{#1}%
\providecommand \citenamefont [1]{#1}%
\providecommand \href@noop [0]{\@secondoftwo}%
\providecommand \href [0]{\begingroup \@sanitize@url \@href}%
\providecommand \@href[1]{\@@startlink{#1}\@@href}%
\providecommand \@@href[1]{\endgroup#1\@@endlink}%
\providecommand \@sanitize@url [0]{\catcode `\\12\catcode `\$12\catcode
  `\&12\catcode `\#12\catcode `\^12\catcode `\_12\catcode `\%12\relax}%
\providecommand \@@startlink[1]{}%
\providecommand \@@endlink[0]{}%
\providecommand \url  [0]{\begingroup\@sanitize@url \@url }%
\providecommand \@url [1]{\endgroup\@href {#1}{\urlprefix }}%
\providecommand \urlprefix  [0]{URL }%
\providecommand \Eprint [0]{\href }%
\providecommand \doibase [0]{http://dx.doi.org/}%
\providecommand \selectlanguage [0]{\@gobble}%
\providecommand \bibinfo  [0]{\@secondoftwo}%
\providecommand \bibfield  [0]{\@secondoftwo}%
\providecommand \translation [1]{[#1]}%
\providecommand \BibitemOpen [0]{}%
\providecommand \bibitemStop [0]{}%
\providecommand \bibitemNoStop [0]{.\EOS\space}%
\providecommand \EOS [0]{\spacefactor3000\relax}%
\providecommand \BibitemShut  [1]{\csname bibitem#1\endcsname}%
\let\auto@bib@innerbib\@empty
\bibitem [{\citenamefont {Dragan}\ \emph
  {et~al.}(2013{\natexlab{a}})\citenamefont {Dragan}, \citenamefont {Doukas},\
  and\ \citenamefont {Mart\'{\i}n-Mart\'{\i}nez}}]{Drago1}%
  \BibitemOpen
  \bibfield  {author} {\bibinfo {author} {\bibfnamefont {A.}~\bibnamefont
  {Dragan}}, \bibinfo {author} {\bibfnamefont {J.}~\bibnamefont {Doukas}}, \
  and\ \bibinfo {author} {\bibfnamefont {E.}~\bibnamefont
  {Mart\'{\i}n-Mart\'{\i}nez}},\ }\href {\doibase 10.1103/PhysRevA.87.052326}
  {\bibfield  {journal} {\bibinfo  {journal} {Phys. Rev. A}\ }\textbf {\bibinfo
  {volume} {87}},\ \bibinfo {pages} {052326} (\bibinfo {year}
  {2013}{\natexlab{a}})}\BibitemShut {NoStop}%
\bibitem [{\citenamefont {Dragan}\ \emph
  {et~al.}(2013{\natexlab{b}})\citenamefont {Dragan}, \citenamefont {Doukas},
  \citenamefont {Mart{\'\i}n-Mart{\'\i}nez},\ and\ \citenamefont
  {Bruschi}}]{Drago2}%
  \BibitemOpen
  \bibfield  {author} {\bibinfo {author} {\bibfnamefont {A.}~\bibnamefont
  {Dragan}}, \bibinfo {author} {\bibfnamefont {J.}~\bibnamefont {Doukas}},
  \bibinfo {author} {\bibfnamefont {E.}~\bibnamefont
  {Mart{\'\i}n-Mart{\'\i}nez}}, \ and\ \bibinfo {author} {\bibfnamefont
  {D.~E.}\ \bibnamefont {Bruschi}},\ }\href
  {http://stacks.iop.org/0264-9381/30/i=23/a=235006} {\bibfield  {journal}
  {\bibinfo  {journal} {Classical and Quantum Gravity}\ }\textbf {\bibinfo
  {volume} {30}},\ \bibinfo {pages} {235006} (\bibinfo {year}
  {2013}{\natexlab{b}})}\BibitemShut {NoStop}%
\bibitem [{\citenamefont {Lin}(2014)}]{Lin2014773}%
  \BibitemOpen
  \bibfield  {author} {\bibinfo {author} {\bibfnamefont {S.-Y.}\ \bibnamefont
  {Lin}},\ }\href {\doibase http://dx.doi.org/10.1016/j.aop.2014.08.018}
  {\bibfield  {journal} {\bibinfo  {journal} {Ann. Phys.}\ }\textbf {\bibinfo
  {volume} {351}},\ \bibinfo {pages} {773 } (\bibinfo {year}
  {2014})}\BibitemShut {NoStop}%
\bibitem [{\citenamefont {Unruh}(1976)}]{unruh_notes_1976}%
  \BibitemOpen
  \bibfield  {author} {\bibinfo {author} {\bibfnamefont {W.~G.}\ \bibnamefont
  {Unruh}},\ }\href {\doibase 10.1103/PhysRevD.14.870} {\bibfield  {journal}
  {\bibinfo  {journal} {Phys. Rev. D}\ }\textbf {\bibinfo {volume} {14}},\
  \bibinfo {pages} {870} (\bibinfo {year} {1976})}\BibitemShut {NoStop}%
\bibitem [{\citenamefont {DeWitt}(1979)}]{DeWitts}%
  \BibitemOpen
  \bibfield  {author} {\bibinfo {author} {\bibfnamefont {B.}~\bibnamefont
  {DeWitt}},\ }in\ \href@noop {} {\emph {\bibinfo {booktitle} {General
  Relativity: An Einstein Centenary Survey}}},\ \bibinfo {editor} {edited by\
  \bibinfo {editor} {\bibfnamefont {S.~W.}\ \bibnamefont {Hawking}}\ and\
  \bibinfo {editor} {\bibfnamefont {W.}~\bibnamefont {Israel}}}\ (\bibinfo
  {publisher} {Cambridge University Press},\ \bibinfo {address} {Cambridge},\
  \bibinfo {year} {1979})\BibitemShut {NoStop}%
\bibitem [{\citenamefont {Lin}\ and\ \citenamefont {Hu}(2007)}]{BeiLok}%
  \BibitemOpen
  \bibfield  {author} {\bibinfo {author} {\bibfnamefont {S.-Y.}\ \bibnamefont
  {Lin}}\ and\ \bibinfo {author} {\bibfnamefont {B.~L.}\ \bibnamefont {Hu}},\
  }\href {\doibase 10.1103/PhysRevD.76.064008} {\bibfield  {journal} {\bibinfo
  {journal} {Phys. Rev. D}\ }\textbf {\bibinfo {volume} {76}},\ \bibinfo
  {pages} {064008} (\bibinfo {year} {2007})}\BibitemShut {NoStop}%
\bibitem [{\citenamefont {Brown}\ \emph {et~al.}(2013)\citenamefont {Brown},
  \citenamefont {Mart\'{i}n-Mart\'{i}nez}, \citenamefont {Menicucci},\ and\
  \citenamefont {Mann}}]{Brown2012}%
  \BibitemOpen
  \bibfield  {author} {\bibinfo {author} {\bibfnamefont {E.~G.}\ \bibnamefont
  {Brown}}, \bibinfo {author} {\bibfnamefont {E.}~\bibnamefont
  {Mart\'{i}n-Mart\'{i}nez}}, \bibinfo {author} {\bibfnamefont {N.~C.}\
  \bibnamefont {Menicucci}}, \ and\ \bibinfo {author} {\bibfnamefont {R.~B.}\
  \bibnamefont {Mann}},\ }\href {\doibase 10.1103/PhysRevD.87.084062}
  {\bibfield  {journal} {\bibinfo  {journal} {Phys. Rev. D}\ }\textbf {\bibinfo
  {volume} {87}},\ \bibinfo {pages} {084062} (\bibinfo {year}
  {2013})}\BibitemShut {NoStop}%
\bibitem [{\citenamefont {Bruschi}\ \emph {et~al.}(2013)\citenamefont
  {Bruschi}, \citenamefont {Lee},\ and\ \citenamefont
  {Fuentes}}]{Fuenetesevolution}%
  \BibitemOpen
  \bibfield  {author} {\bibinfo {author} {\bibfnamefont {D.~E.}\ \bibnamefont
  {Bruschi}}, \bibinfo {author} {\bibfnamefont {A.~R.}\ \bibnamefont {Lee}}, \
  and\ \bibinfo {author} {\bibfnamefont {I.}~\bibnamefont {Fuentes}},\
  }\href@noop {} {\bibfield  {journal} {\bibinfo  {journal} {J. Phys. A: Math.
  Theor.}\ }\textbf {\bibinfo {volume} {46}},\ \bibinfo {pages} {165303}
  (\bibinfo {year} {2013})}\BibitemShut {NoStop}%
\bibitem [{\citenamefont {Crispino}\ \emph {et~al.}(2008)\citenamefont
  {Crispino}, \citenamefont {Higuchi},\ and\ \citenamefont
  {Matsas}}]{Crispino}%
  \BibitemOpen
  \bibfield  {author} {\bibinfo {author} {\bibfnamefont {L.~C.~B.}\
  \bibnamefont {Crispino}}, \bibinfo {author} {\bibfnamefont {A.}~\bibnamefont
  {Higuchi}}, \ and\ \bibinfo {author} {\bibfnamefont {G.~E.~A.}\ \bibnamefont
  {Matsas}},\ }\href@noop {} {\bibfield  {journal} {\bibinfo  {journal} {Rev.
  Mod. Phys.}\ }\textbf {\bibinfo {volume} {80}},\ \bibinfo {pages} {787}
  (\bibinfo {year} {2008})}\BibitemShut {NoStop}%
\bibitem [{\citenamefont {Takagi}(1986)}]{Takagi}%
  \BibitemOpen
  \bibfield  {author} {\bibinfo {author} {\bibfnamefont {S.}~\bibnamefont
  {Takagi}},\ }\href@noop {} {\bibfield  {journal} {\bibinfo  {journal} {Prog.
  Theor. Phys. Suppl.}\ }\textbf {\bibinfo {volume} {88}},\ \bibinfo {pages}
  {1} (\bibinfo {year} {1986})}\BibitemShut {NoStop}%
\bibitem [{\citenamefont {Candelas}\ and\ \citenamefont
  {Sciama}(1977)}]{candelas_irreversible_1977}%
  \BibitemOpen
  \bibfield  {author} {\bibinfo {author} {\bibfnamefont {P.}~\bibnamefont
  {Candelas}}\ and\ \bibinfo {author} {\bibfnamefont {D.~W.}\ \bibnamefont
  {Sciama}},\ }\href {\doibase 10.1103/PhysRevLett.38.1372} {\bibfield
  {journal} {\bibinfo  {journal} {Phys. Rev. Lett.}\ }\textbf {\bibinfo
  {volume} {38}},\ \bibinfo {pages} {1372} (\bibinfo {year}
  {1977})}\BibitemShut {NoStop}%
\bibitem [{\citenamefont {Scully}\ and\ \citenamefont
  {Zubairy}(1997)}]{scullybook}%
  \BibitemOpen
  \bibfield  {author} {\bibinfo {author} {\bibfnamefont {M.~O.}\ \bibnamefont
  {Scully}}\ and\ \bibinfo {author} {\bibfnamefont {M.~S.}\ \bibnamefont
  {Zubairy}},\ }\href@noop {} {\emph {\bibinfo {title} {Quantum Optics}}}\
  (\bibinfo  {publisher} {Cambridge University Press},\ \bibinfo {year}
  {1997})\BibitemShut {NoStop}%
\bibitem [{\citenamefont {Wallraff}\ \emph {et~al.}(2004)\citenamefont
  {Wallraff}, \citenamefont {Schuster}, \citenamefont {Blais}, \citenamefont
  {Frunzio}, \citenamefont {Huang}, \citenamefont {Majer}, \citenamefont
  {Kumar}, \citenamefont {Girvin},\ and\ \citenamefont
  {Schoelkopf}}]{Wallraff:2004aa}%
  \BibitemOpen
  \bibfield  {author} {\bibinfo {author} {\bibfnamefont {A.}~\bibnamefont
  {Wallraff}}, \bibinfo {author} {\bibfnamefont {D.~I.}\ \bibnamefont
  {Schuster}}, \bibinfo {author} {\bibfnamefont {A.}~\bibnamefont {Blais}},
  \bibinfo {author} {\bibfnamefont {L.}~\bibnamefont {Frunzio}}, \bibinfo
  {author} {\bibfnamefont {R.~S.}\ \bibnamefont {Huang}}, \bibinfo {author}
  {\bibfnamefont {J.}~\bibnamefont {Majer}}, \bibinfo {author} {\bibfnamefont
  {S.}~\bibnamefont {Kumar}}, \bibinfo {author} {\bibfnamefont {S.~M.}\
  \bibnamefont {Girvin}}, \ and\ \bibinfo {author} {\bibfnamefont {R.~J.}\
  \bibnamefont {Schoelkopf}},\ }\href {http://dx.doi.org/10.1038/nature02851}
  {\bibfield  {journal} {\bibinfo  {journal} {Nature}\ }\textbf {\bibinfo
  {volume} {431}},\ \bibinfo {pages} {162} (\bibinfo {year}
  {2004})}\BibitemShut {NoStop}%
\bibitem [{\citenamefont {Mart\'{i}n-Mart\'{i}nez}\ \emph
  {et~al.}(2013{\natexlab{a}})\citenamefont {Mart\'{i}n-Mart\'{i}nez},
  \citenamefont {Montero},\ and\ \citenamefont {del Rey}}]{Wavepackets}%
  \BibitemOpen
  \bibfield  {author} {\bibinfo {author} {\bibfnamefont {E.}~\bibnamefont
  {Mart\'{i}n-Mart\'{i}nez}}, \bibinfo {author} {\bibfnamefont
  {M.}~\bibnamefont {Montero}}, \ and\ \bibinfo {author} {\bibfnamefont
  {M.}~\bibnamefont {del Rey}},\ }\href {\doibase 10.1103/PhysRevD.87.064038}
  {\bibfield  {journal} {\bibinfo  {journal} {Phys. Rev. D}\ }\textbf {\bibinfo
  {volume} {87}},\ \bibinfo {pages} {064038} (\bibinfo {year}
  {2013}{\natexlab{a}})}\BibitemShut {NoStop}%
\bibitem [{\citenamefont {Alhambra}\ \emph {et~al.}(2014)\citenamefont
  {Alhambra}, \citenamefont {Kempf},\ and\ \citenamefont
  {Mart\'in-Mart\'inez}}]{Alvaro}%
  \BibitemOpen
  \bibfield  {author} {\bibinfo {author} {\bibfnamefont {A.~M.}\ \bibnamefont
  {Alhambra}}, \bibinfo {author} {\bibfnamefont {A.}~\bibnamefont {Kempf}}, \
  and\ \bibinfo {author} {\bibfnamefont {E.}~\bibnamefont
  {Mart\'in-Mart\'inez}},\ }\href {\doibase 10.1103/PhysRevA.89.033835}
  {\bibfield  {journal} {\bibinfo  {journal} {Phys. Rev. A}\ }\textbf {\bibinfo
  {volume} {89}},\ \bibinfo {pages} {033835} (\bibinfo {year}
  {2014})}\BibitemShut {NoStop}%
\bibitem [{\citenamefont {Mart\'{i}n-Mart\'{i}nez}\ \emph
  {et~al.}(2013{\natexlab{b}})\citenamefont {Mart\'{i}n-Mart\'{i}nez},
  \citenamefont {Aasen},\ and\ \citenamefont {Kempf}}]{AasenPRL}%
  \BibitemOpen
  \bibfield  {author} {\bibinfo {author} {\bibfnamefont {E.}~\bibnamefont
  {Mart\'{i}n-Mart\'{i}nez}}, \bibinfo {author} {\bibfnamefont
  {D.}~\bibnamefont {Aasen}}, \ and\ \bibinfo {author} {\bibfnamefont
  {A.}~\bibnamefont {Kempf}},\ }\href {\doibase 10.1103/PhysRevLett.110.160501}
  {\bibfield  {journal} {\bibinfo  {journal} {Phys. Rev. Lett.}\ }\textbf
  {\bibinfo {volume} {110}},\ \bibinfo {pages} {160501} (\bibinfo {year}
  {2013}{\natexlab{b}})}\BibitemShut {NoStop}%
\bibitem [{\citenamefont {Mart\'{i}n-Mart\'{i}nez}\ and\ \citenamefont
  {Sutherland}(2014)}]{Chris}%
  \BibitemOpen
  \bibfield  {author} {\bibinfo {author} {\bibfnamefont {E.}~\bibnamefont
  {Mart\'{i}n-Mart\'{i}nez}}\ and\ \bibinfo {author} {\bibfnamefont
  {C.}~\bibnamefont {Sutherland}},\ }\href {\doibase
  http://dx.doi.org/10.1016/j.physletb.2014.10.038} {\bibfield  {journal}
  {\bibinfo  {journal} {Phys. Lett. B}\ }\textbf {\bibinfo {volume} {739}},\
  \bibinfo {pages} {74 } (\bibinfo {year} {2014})}\BibitemShut {NoStop}%
\bibitem [{\citenamefont {Cliche}\ and\ \citenamefont
  {Kempf}(2010)}]{mathieuachim1}%
  \BibitemOpen
  \bibfield  {author} {\bibinfo {author} {\bibfnamefont {M.}~\bibnamefont
  {Cliche}}\ and\ \bibinfo {author} {\bibfnamefont {A.}~\bibnamefont {Kempf}},\
  }\href {\doibase 10.1103/PhysRevA.81.012330} {\bibfield  {journal} {\bibinfo
  {journal} {Phys. Rev. A}\ }\textbf {\bibinfo {volume} {81}},\ \bibinfo
  {pages} {012330} (\bibinfo {year} {2010})}\BibitemShut {NoStop}%
\bibitem [{\citenamefont {Jonsson}\ \emph {et~al.}(2014)\citenamefont
  {Jonsson}, \citenamefont {Mart\'{i}n-Mart\'{i}nez},\ and\ \citenamefont
  {Kempf}}]{Robort}%
  \BibitemOpen
  \bibfield  {author} {\bibinfo {author} {\bibfnamefont {R.~H.}\ \bibnamefont
  {Jonsson}}, \bibinfo {author} {\bibfnamefont {E.}~\bibnamefont
  {Mart\'{i}n-Mart\'{i}nez}}, \ and\ \bibinfo {author} {\bibfnamefont
  {A.}~\bibnamefont {Kempf}},\ }\href {\doibase 10.1103/PhysRevA.89.022330}
  {\bibfield  {journal} {\bibinfo  {journal} {Phys. Rev. A}\ }\textbf {\bibinfo
  {volume} {89}},\ \bibinfo {pages} {022330} (\bibinfo {year}
  {2014})}\BibitemShut {NoStop}%
\bibitem [{\citenamefont {Jonsson}\ \emph {et~al.}(2015)\citenamefont
  {Jonsson}, \citenamefont {Mart\'{i}n-Mart\'{i}nez},\ and\ \citenamefont
  {Kempf}}]{Robort2}%
  \BibitemOpen
  \bibfield  {author} {\bibinfo {author} {\bibfnamefont {R.~H.}\ \bibnamefont
  {Jonsson}}, \bibinfo {author} {\bibfnamefont {E.}~\bibnamefont
  {Mart\'{i}n-Mart\'{i}nez}}, \ and\ \bibinfo {author} {\bibfnamefont
  {A.}~\bibnamefont {Kempf}},\ }\href {\doibase 10.1103/PhysRevLett.114.110505}
  {\bibfield  {journal} {\bibinfo  {journal} {Phys. Rev. Lett.}\ }\textbf
  {\bibinfo {volume} {114}},\ \bibinfo {pages} {110505} (\bibinfo {year}
  {2015})}\BibitemShut {NoStop}%
\bibitem [{\citenamefont {Gibbons}\ and\ \citenamefont
  {Hawking}(1977)}]{Gibbons1977}%
  \BibitemOpen
  \bibfield  {author} {\bibinfo {author} {\bibfnamefont {G.~W.}\ \bibnamefont
  {Gibbons}}\ and\ \bibinfo {author} {\bibfnamefont {S.~W.}\ \bibnamefont
  {Hawking}},\ }\href {\doibase 10.1103/PhysRevD.15.2738} {\bibfield  {journal}
  {\bibinfo  {journal} {Phys. Rev. D}\ }\textbf {\bibinfo {volume} {15}},\
  \bibinfo {pages} {2738} (\bibinfo {year} {1977})}\BibitemShut {NoStop}%
\bibitem [{\citenamefont {Garay}\ \emph {et~al.}(2014)\citenamefont {Garay},
  \citenamefont {Mart\'in-Benito},\ and\ \citenamefont
  {Mart\'in-Mart\'inez}}]{QuanG}%
  \BibitemOpen
  \bibfield  {author} {\bibinfo {author} {\bibfnamefont {L.~J.}\ \bibnamefont
  {Garay}}, \bibinfo {author} {\bibfnamefont {M.}~\bibnamefont
  {Mart\'in-Benito}}, \ and\ \bibinfo {author} {\bibfnamefont {E.}~\bibnamefont
  {Mart\'in-Mart\'inez}},\ }\href {\doibase 10.1103/PhysRevD.89.043510}
  {\bibfield  {journal} {\bibinfo  {journal} {Phys. Rev. D}\ }\textbf {\bibinfo
  {volume} {89}},\ \bibinfo {pages} {043510} (\bibinfo {year}
  {2014})}\BibitemShut {NoStop}%
\bibitem [{\citenamefont {Mart\'{i}n-Mart\'{i}nez}\ and\ \citenamefont
  {Menicucci}(2012)}]{cosmoq}%
  \BibitemOpen
  \bibfield  {author} {\bibinfo {author} {\bibfnamefont {E.}~\bibnamefont
  {Mart\'{i}n-Mart\'{i}nez}}\ and\ \bibinfo {author} {\bibfnamefont {N.~C.}\
  \bibnamefont {Menicucci}},\ }\href {\doibase 10.1088/0264-9381/29/22/224003}
  {\bibfield  {journal} {\bibinfo  {journal} {Classical and Quantum Gravity}\
  }\textbf {\bibinfo {volume} {29}},\ \bibinfo {pages} {224003} (\bibinfo
  {year} {2012})}\BibitemShut {NoStop}%
\bibitem [{\citenamefont {Valentini}(1991)}]{Valentini1991}%
  \BibitemOpen
  \bibfield  {author} {\bibinfo {author} {\bibfnamefont {A.}~\bibnamefont
  {Valentini}},\ }\href {\doibase
  http://dx.doi.org/10.1016/0375-9601(91)90952-5} {\bibfield  {journal}
  {\bibinfo  {journal} {Physics Letters A}\ }\textbf {\bibinfo {volume}
  {153}},\ \bibinfo {pages} {321 } (\bibinfo {year} {1991})}\BibitemShut
  {NoStop}%
\bibitem [{\citenamefont {Reznik}(2003)}]{Reznik2003}%
  \BibitemOpen
  \bibfield  {author} {\bibinfo {author} {\bibfnamefont {B.}~\bibnamefont
  {Reznik}},\ }\href {\doibase 10.1023/A:1022875910744} {\bibfield  {journal}
  {\bibinfo  {journal} {Foundations of Physics}\ }\textbf {\bibinfo {volume}
  {33}},\ \bibinfo {pages} {167} (\bibinfo {year} {2003})}\BibitemShut
  {NoStop}%
\bibitem [{\citenamefont {Reznik}\ \emph {et~al.}(2005)\citenamefont {Reznik},
  \citenamefont {Retzker},\ and\ \citenamefont {Silman}}]{Reznik1}%
  \BibitemOpen
  \bibfield  {author} {\bibinfo {author} {\bibfnamefont {B.}~\bibnamefont
  {Reznik}}, \bibinfo {author} {\bibfnamefont {A.}~\bibnamefont {Retzker}}, \
  and\ \bibinfo {author} {\bibfnamefont {J.}~\bibnamefont {Silman}},\ }\href
  {http://link.aps.org/abstract/PRA/v71/e042104} {\bibfield  {journal}
  {\bibinfo  {journal} {Phys. Rev. A}\ }\textbf {\bibinfo {volume} {71}},\
  \bibinfo {eid} {042104} (\bibinfo {year} {2005})}\BibitemShut {NoStop}%
\bibitem [{\citenamefont {Ver~Steeg}\ and\ \citenamefont
  {Menicucci}(2009)}]{VerSteeg2009}%
  \BibitemOpen
  \bibfield  {author} {\bibinfo {author} {\bibfnamefont {G.}~\bibnamefont
  {Ver~Steeg}}\ and\ \bibinfo {author} {\bibfnamefont {N.~C.}\ \bibnamefont
  {Menicucci}},\ }\href {\doibase 10.1103/PhysRevD.79.044027} {\bibfield
  {journal} {\bibinfo  {journal} {Phys. Rev. D}\ }\textbf {\bibinfo {volume}
  {79}},\ \bibinfo {pages} {044027} (\bibinfo {year} {2009})}\BibitemShut
  {NoStop}%
\bibitem [{\citenamefont {Olson}\ and\ \citenamefont
  {Ralph}(2011)}]{Olson2011}%
  \BibitemOpen
  \bibfield  {author} {\bibinfo {author} {\bibfnamefont {S.~J.}\ \bibnamefont
  {Olson}}\ and\ \bibinfo {author} {\bibfnamefont {T.~C.}\ \bibnamefont
  {Ralph}},\ }\href {\doibase 10.1103/PhysRevLett.106.110404} {\bibfield
  {journal} {\bibinfo  {journal} {Phys. Rev. Lett.}\ }\textbf {\bibinfo
  {volume} {106}},\ \bibinfo {pages} {110404} (\bibinfo {year}
  {2011})}\BibitemShut {NoStop}%
\bibitem [{\citenamefont {Salton}\ \emph {et~al.}(2015)\citenamefont {Salton},
  \citenamefont {Mann},\ and\ \citenamefont {Menicucci}}]{Salton:2014jaa}%
  \BibitemOpen
  \bibfield  {author} {\bibinfo {author} {\bibfnamefont {G.}~\bibnamefont
  {Salton}}, \bibinfo {author} {\bibfnamefont {R.~B.}\ \bibnamefont {Mann}}, \
  and\ \bibinfo {author} {\bibfnamefont {N.~C.}\ \bibnamefont {Menicucci}},\
  }\href {\doibase 10.1088/1367-2630/17/3/035001} {\bibfield  {journal}
  {\bibinfo  {journal} {New J. Phys.}\ }\textbf {\bibinfo {volume} {17}},\
  \bibinfo {pages} {035001} (\bibinfo {year} {2015})}\BibitemShut {NoStop}%
\bibitem [{\citenamefont {Pozas-Kerstjens}\ and\ \citenamefont
  {Mart\'{\i}n-Mart\'{\i}nez}(2015)}]{Pozas-Kerstjens:2015}%
  \BibitemOpen
  \bibfield  {author} {\bibinfo {author} {\bibfnamefont {A.}~\bibnamefont
  {Pozas-Kerstjens}}\ and\ \bibinfo {author} {\bibfnamefont {E.}~\bibnamefont
  {Mart\'{\i}n-Mart\'{\i}nez}},\ }\href {\doibase 10.1103/PhysRevD.92.064042}
  {\bibfield  {journal} {\bibinfo  {journal} {Phys. Rev. D}\ }\textbf {\bibinfo
  {volume} {92}},\ \bibinfo {pages} {064042} (\bibinfo {year}
  {2015})}\BibitemShut {NoStop}%
\bibitem [{\citenamefont {Mart\'{i}n-Mart\'{i}nez}\ \emph
  {et~al.}(2015)\citenamefont {Mart\'{i}n-Mart\'{i}nez}, \citenamefont
  {Smith},\ and\ \citenamefont {Terno}}]{Martin-MartinezSmithTerno}%
  \BibitemOpen
  \bibfield  {author} {\bibinfo {author} {\bibfnamefont {E.}~\bibnamefont
  {Mart\'{i}n-Mart\'{i}nez}}, \bibinfo {author} {\bibfnamefont
  {A.}~\bibnamefont {Smith}}, \ and\ \bibinfo {author} {\bibfnamefont
  {D.}~\bibnamefont {Terno}},\ }\href {http://arxiv.org/abs/1507.02688} {\
  (\bibinfo {year} {2015})},\ \Eprint {http://arxiv.org/abs/arXiv:1507.02688}
  {arXiv:1507.02688} \BibitemShut {NoStop}%
\bibitem [{\citenamefont {Mart\'{i}n-Mart\'{i}nez}\ \emph
  {et~al.}(2013{\natexlab{c}})\citenamefont {Mart\'{i}n-Mart\'{i}nez},
  \citenamefont {Brown}, \citenamefont {Donnelly},\ and\ \citenamefont
  {Kempf}}]{Farming}%
  \BibitemOpen
  \bibfield  {author} {\bibinfo {author} {\bibfnamefont {E.}~\bibnamefont
  {Mart\'{i}n-Mart\'{i}nez}}, \bibinfo {author} {\bibfnamefont {E.~G.}\
  \bibnamefont {Brown}}, \bibinfo {author} {\bibfnamefont {W.}~\bibnamefont
  {Donnelly}}, \ and\ \bibinfo {author} {\bibfnamefont {A.}~\bibnamefont
  {Kempf}},\ }\href {\doibase 10.1103/PhysRevA.88.052310} {\bibfield  {journal}
  {\bibinfo  {journal} {Phys. Rev. A}\ }\textbf {\bibinfo {volume} {88}},\
  \bibinfo {pages} {052310} (\bibinfo {year} {2013}{\natexlab{c}})}\BibitemShut
  {NoStop}%
\bibitem [{\citenamefont {Brown}\ \emph {et~al.}(2014)\citenamefont {Brown},
  \citenamefont {Donnelly}, \citenamefont {Kempf}, \citenamefont {Mann},
  \citenamefont {Mart{\'\i}n-Mart{\'\i}nez},\ and\ \citenamefont
  {Menicucci}}]{Brown:2014en}%
  \BibitemOpen
  \bibfield  {author} {\bibinfo {author} {\bibfnamefont {E.~G.}\ \bibnamefont
  {Brown}}, \bibinfo {author} {\bibfnamefont {W.}~\bibnamefont {Donnelly}},
  \bibinfo {author} {\bibfnamefont {A.}~\bibnamefont {Kempf}}, \bibinfo
  {author} {\bibfnamefont {R.~B.}\ \bibnamefont {Mann}}, \bibinfo {author}
  {\bibfnamefont {E.}~\bibnamefont {Mart{\'\i}n-Mart{\'\i}nez}}, \ and\
  \bibinfo {author} {\bibfnamefont {N.~C.}\ \bibnamefont {Menicucci}},\ }\href
  {\doibase 10.1088/1367-2630/16/10/105020} {\bibfield  {journal} {\bibinfo
  {journal} {New J. Phys.}\ }\textbf {\bibinfo {volume} {16}},\ \bibinfo
  {pages} {105020} (\bibinfo {year} {2014})}\BibitemShut {NoStop}%
\bibitem [{\citenamefont {Summers}\ and\ \citenamefont
  {Werner}(1985)}]{Alegbra1}%
  \BibitemOpen
  \bibfield  {author} {\bibinfo {author} {\bibfnamefont {S.~J.}\ \bibnamefont
  {Summers}}\ and\ \bibinfo {author} {\bibfnamefont {R.~F.}\ \bibnamefont
  {Werner}},\ }\href@noop {} {\bibfield  {journal} {\bibinfo  {journal} {Phys.
  Lett. A}\ }\textbf {\bibinfo {volume} {110}},\ \bibinfo {pages} {257}
  (\bibinfo {year} {1985})}\BibitemShut {NoStop}%
\bibitem [{\citenamefont {Summers}\ and\ \citenamefont
  {Werner}(1987)}]{Alegbra2}%
  \BibitemOpen
  \bibfield  {author} {\bibinfo {author} {\bibfnamefont {S.~J.}\ \bibnamefont
  {Summers}}\ and\ \bibinfo {author} {\bibfnamefont {R.~F.}\ \bibnamefont
  {Werner}},\ }\href@noop {} {\bibfield  {journal} {\bibinfo  {journal} {J.
  Math. Phys.}\ }\textbf {\bibinfo {volume} {28}},\ \bibinfo {pages} {2440}
  (\bibinfo {year} {1987})}\BibitemShut {NoStop}%
\bibitem [{\citenamefont {Benincasa}\ \emph {et~al.}(2014)\citenamefont
  {Benincasa}, \citenamefont {Borsten}, \citenamefont {Buck},\ and\
  \citenamefont {Dowker}}]{BenicasaBorstenBuckDowker}%
  \BibitemOpen
  \bibfield  {author} {\bibinfo {author} {\bibfnamefont {D.~M.~T.}\
  \bibnamefont {Benincasa}}, \bibinfo {author} {\bibfnamefont {L.}~\bibnamefont
  {Borsten}}, \bibinfo {author} {\bibfnamefont {M.}~\bibnamefont {Buck}}, \
  and\ \bibinfo {author} {\bibfnamefont {F.}~\bibnamefont {Dowker}},\ }\href
  {http://stacks.iop.org/0264-9381/31/i=7/a=075007} {\bibfield  {journal}
  {\bibinfo  {journal} {Class. Quant. Grav.}\ }\textbf {\bibinfo {volume}
  {31}},\ \bibinfo {pages} {075007} (\bibinfo {year} {2014})}\BibitemShut
  {NoStop}%
\bibitem [{\citenamefont {Langlois}(2006)}]{Langlois2006}%
  \BibitemOpen
  \bibfield  {author} {\bibinfo {author} {\bibfnamefont {P.}~\bibnamefont
  {Langlois}},\ }\href {\doibase 10.1016/j.aop.2006.01.013} {\bibfield
  {journal} {\bibinfo  {journal} {Annals Phys.}\ }\textbf {\bibinfo {volume}
  {321}},\ \bibinfo {pages} {2027 } (\bibinfo {year} {2006})}\BibitemShut
  {NoStop}%
\bibitem [{\citenamefont {Louko}\ and\ \citenamefont {Satz}(2006)}]{Satz2006}%
  \BibitemOpen
  \bibfield  {author} {\bibinfo {author} {\bibfnamefont {J.}~\bibnamefont
  {Louko}}\ and\ \bibinfo {author} {\bibfnamefont {A.}~\bibnamefont {Satz}},\
  }\href@noop {} {\bibfield  {journal} {\bibinfo  {journal} {Class. Quant.
  Grav.}\ }\textbf {\bibinfo {volume} {23}},\ \bibinfo {pages} {6321} (\bibinfo
  {year} {2006})}\BibitemShut {NoStop}%
\bibitem [{\citenamefont {Schlicht}(2004)}]{Schlicht}%
  \BibitemOpen
  \bibfield  {author} {\bibinfo {author} {\bibfnamefont {S.}~\bibnamefont
  {Schlicht}},\ }\href@noop {} {\bibfield  {journal} {\bibinfo  {journal}
  {Class. Quant. Grav.}\ }\textbf {\bibinfo {volume} {21}},\ \bibinfo {pages}
  {4647} (\bibinfo {year} {2004})}\BibitemShut {NoStop}%
\bibitem [{\citenamefont {Power}\ and\ \citenamefont
  {Thirunamachandran}(1983)}]{Tiramisu}%
  \BibitemOpen
  \bibfield  {author} {\bibinfo {author} {\bibfnamefont {E.~A.}\ \bibnamefont
  {Power}}\ and\ \bibinfo {author} {\bibfnamefont {T.}~\bibnamefont
  {Thirunamachandran}},\ }\href {\doibase 10.1103/PhysRevA.28.2663} {\bibfield
  {journal} {\bibinfo  {journal} {Phys. Rev. A}\ }\textbf {\bibinfo {volume}
  {28}},\ \bibinfo {pages} {2663} (\bibinfo {year} {1983})}\BibitemShut
  {NoStop}%
\bibitem [{\citenamefont {Milonni}\ \emph {et~al.}(1995)\citenamefont
  {Milonni}, \citenamefont {James},\ and\ \citenamefont {Fearn}}]{Milo}%
  \BibitemOpen
  \bibfield  {author} {\bibinfo {author} {\bibfnamefont {P.~W.}\ \bibnamefont
  {Milonni}}, \bibinfo {author} {\bibfnamefont {D.~F.~V.}\ \bibnamefont
  {James}}, \ and\ \bibinfo {author} {\bibfnamefont {H.}~\bibnamefont
  {Fearn}},\ }\href {\doibase 10.1103/PhysRevA.52.1525} {\bibfield  {journal}
  {\bibinfo  {journal} {Phys. Rev. A}\ }\textbf {\bibinfo {volume} {52}},\
  \bibinfo {pages} {1525} (\bibinfo {year} {1995})}\BibitemShut {NoStop}%
\bibitem [{\citenamefont {Dolce}\ \emph {et~al.}(2006)\citenamefont {Dolce},
  \citenamefont {Passante},\ and\ \citenamefont {Persico}}]{Dolce}%
  \BibitemOpen
  \bibfield  {author} {\bibinfo {author} {\bibfnamefont {I.}~\bibnamefont
  {Dolce}}, \bibinfo {author} {\bibfnamefont {R.}~\bibnamefont {Passante}}, \
  and\ \bibinfo {author} {\bibfnamefont {F.}~\bibnamefont {Persico}},\ }\href
  {\doibase http://dx.doi.org/10.1016/j.physleta.2006.02.018} {\bibfield
  {journal} {\bibinfo  {journal} {Phys. Lett. A}\ }\textbf {\bibinfo {volume}
  {355}},\ \bibinfo {pages} {152 } (\bibinfo {year} {2006})}\BibitemShut
  {NoStop}%
\bibitem [{\citenamefont {Plimak}\ and\ \citenamefont
  {Stenholm}(2011)}]{NonRWA}%
  \BibitemOpen
  \bibfield  {author} {\bibinfo {author} {\bibfnamefont {L.~I.}\ \bibnamefont
  {Plimak}}\ and\ \bibinfo {author} {\bibfnamefont {S.~T.}\ \bibnamefont
  {Stenholm}},\ }\href {http://stacks.iop.org/0295-5075/96/i=3/a=34002}
  {\bibfield  {journal} {\bibinfo  {journal} {EPL (Europhysics Letters)}\
  }\textbf {\bibinfo {volume} {96}},\ \bibinfo {pages} {34002} (\bibinfo {year}
  {2011})}\BibitemShut {NoStop}%
\bibitem [{\citenamefont {Wilson}\ \emph {et~al.}(2011)\citenamefont {Wilson},
  \citenamefont {Johansson}, \citenamefont {Pourkabirian}, \citenamefont
  {Simoen}, \citenamefont {Johansson}, \citenamefont {Duty}, \citenamefont
  {Nori},\ and\ \citenamefont {Delsing}}]{DCasimir}%
  \BibitemOpen
  \bibfield  {author} {\bibinfo {author} {\bibfnamefont {C.~M.}\ \bibnamefont
  {Wilson}}, \bibinfo {author} {\bibfnamefont {G.}~\bibnamefont {Johansson}},
  \bibinfo {author} {\bibfnamefont {A.}~\bibnamefont {Pourkabirian}}, \bibinfo
  {author} {\bibfnamefont {M.}~\bibnamefont {Simoen}}, \bibinfo {author}
  {\bibfnamefont {J.~R.}\ \bibnamefont {Johansson}}, \bibinfo {author}
  {\bibfnamefont {T.}~\bibnamefont {Duty}}, \bibinfo {author} {\bibfnamefont
  {F.}~\bibnamefont {Nori}}, \ and\ \bibinfo {author} {\bibfnamefont
  {P.}~\bibnamefont {Delsing}},\ }\href@noop {} {\bibfield  {journal} {\bibinfo
   {journal} {Nature}\ }\textbf {\bibinfo {volume} {479}},\ \bibinfo {pages}
  {376} (\bibinfo {year} {2011})}\BibitemShut {NoStop}%
\bibitem [{\citenamefont {Nation}\ \emph {et~al.}(2009)\citenamefont {Nation},
  \citenamefont {Blencowe}, \citenamefont {Rimberg},\ and\ \citenamefont
  {Buks}}]{supercond}%
  \BibitemOpen
  \bibfield  {author} {\bibinfo {author} {\bibfnamefont {P.~D.}\ \bibnamefont
  {Nation}}, \bibinfo {author} {\bibfnamefont {M.~P.}\ \bibnamefont
  {Blencowe}}, \bibinfo {author} {\bibfnamefont {A.~J.}\ \bibnamefont
  {Rimberg}}, \ and\ \bibinfo {author} {\bibfnamefont {E.}~\bibnamefont
  {Buks}},\ }\href@noop {} {\bibfield  {journal} {\bibinfo  {journal} {Phys.
  Rev. Lett.}\ }\textbf {\bibinfo {volume} {103}},\ \bibinfo {pages} {087004}
  (\bibinfo {year} {2009})}\BibitemShut {NoStop}%
\bibitem [{\citenamefont {Peropadre}\ \emph {et~al.}(2010)\citenamefont
  {Peropadre}, \citenamefont {Forn-D\'iaz}, \citenamefont {Solano},\ and\
  \citenamefont {Garc\'ia-Ripoll}}]{Ultrastrong}%
  \BibitemOpen
  \bibfield  {author} {\bibinfo {author} {\bibfnamefont {B.}~\bibnamefont
  {Peropadre}}, \bibinfo {author} {\bibfnamefont {P.}~\bibnamefont
  {Forn-D\'iaz}}, \bibinfo {author} {\bibfnamefont {E.}~\bibnamefont {Solano}},
  \ and\ \bibinfo {author} {\bibfnamefont {J.~J.}\ \bibnamefont
  {Garc\'ia-Ripoll}},\ }\href {\doibase 10.1103/PhysRevLett.105.023601}
  {\bibfield  {journal} {\bibinfo  {journal} {Phys. Rev. Lett.}\ }\textbf
  {\bibinfo {volume} {105}},\ \bibinfo {pages} {023601} (\bibinfo {year}
  {2010})}\BibitemShut {NoStop}%
\bibitem [{\citenamefont {Dewan}\ and\ \citenamefont
  {Beran}(1959)}]{BellParadox}%
  \BibitemOpen
  \bibfield  {author} {\bibinfo {author} {\bibfnamefont {E.}~\bibnamefont
  {Dewan}}\ and\ \bibinfo {author} {\bibfnamefont {M.}~\bibnamefont {Beran}},\
  }\href {\doibase 10.1119/1.1996214} {\bibfield  {journal} {\bibinfo
  {journal} {Am. J. Phys.}\ }\textbf {\bibinfo {volume} {27}},\ \bibinfo
  {pages} {517} (\bibinfo {year} {1959})}\BibitemShut {NoStop}%
\bibitem [{\citenamefont {Langlois}(2004)}]{Langlois}%
  \BibitemOpen
  \bibfield  {author} {\bibinfo {author} {\bibfnamefont {P.}~\bibnamefont
  {Langlois}},\ }\href@noop {} {\bibfield  {journal} {\bibinfo  {journal}
  {Phys. Rev. D}\ }\textbf {\bibinfo {volume} {70}},\ \bibinfo {pages} {104008}
  (\bibinfo {year} {2004})}\BibitemShut {NoStop}%
\bibitem [{\citenamefont {Mart\'{i}n-Mart\'{i}nez}\ and\ \citenamefont
  {Louko}(2015)}]{JormaEduPRL}%
  \BibitemOpen
  \bibfield  {author} {\bibinfo {author} {\bibfnamefont {E.}~\bibnamefont
  {Mart\'{i}n-Mart\'{i}nez}}\ and\ \bibinfo {author} {\bibfnamefont
  {J.}~\bibnamefont {Louko}},\ }\href {\doibase 10.1103/PhysRevLett.115.031301}
  {\bibfield  {journal} {\bibinfo  {journal} {Phys. Rev. Lett.}\ }\textbf
  {\bibinfo {volume} {115}},\ \bibinfo {pages} {031301} (\bibinfo {year}
  {2015})}\BibitemShut {NoStop}%
\bibitem [{\citenamefont {Louko}\ and\ \citenamefont {Satz}(2008)}]{Louko2008}%
  \BibitemOpen
  \bibfield  {author} {\bibinfo {author} {\bibfnamefont {J.}~\bibnamefont
  {Louko}}\ and\ \bibinfo {author} {\bibfnamefont {A.}~\bibnamefont {Satz}},\
  }\href@noop {} {\bibfield  {journal} {\bibinfo  {journal} {Class. Quant.
  Grav.}\ }\textbf {\bibinfo {volume} {25}},\ \bibinfo {pages} {055012}
  (\bibinfo {year} {2008})}\BibitemShut {NoStop}%
\bibitem [{\citenamefont {McLenaghan}(1974)}]{McLenaghan}%
  \BibitemOpen
  \bibfield  {author} {\bibinfo {author} {\bibfnamefont {R.}~\bibnamefont
  {McLenaghan}},\ }\href@noop {} {\bibfield  {journal} {\bibinfo  {journal}
  {Ann. Inst. H. Poincare}\ }\textbf {\bibinfo {volume} {20}},\ \bibinfo
  {pages} {153} (\bibinfo {year} {1974})}\BibitemShut {NoStop}%
\bibitem [{\citenamefont {Czapor}\ and\ \citenamefont
  {McLenaghan}(2008)}]{czapor}%
  \BibitemOpen
  \bibfield  {author} {\bibinfo {author} {\bibfnamefont {S.}~\bibnamefont
  {Czapor}}\ and\ \bibinfo {author} {\bibfnamefont {R.}~\bibnamefont
  {McLenaghan}},\ }\href
  {http://www.actaphys.uj.edu.pl/_old/sup1/pdf/s1p0055.pdf} {\bibfield
  {journal} {\bibinfo  {journal} {Acta. Phys. Pol. B Proc. Suppl. 1}\ }\textbf
  {\bibinfo {volume} {1}},\ \bibinfo {pages} {55} (\bibinfo {year}
  {2008})}\BibitemShut {NoStop}%
\bibitem [{\citenamefont {Blasco}\ \emph {et~al.}(2015)\citenamefont {Blasco},
  \citenamefont {Garay}, \citenamefont {Mart\'in-Benito},\ and\ \citenamefont
  {Mart\'{i}n-Mart\'{i}nez}}]{Blasco:2015eya}%
  \BibitemOpen
  \bibfield  {author} {\bibinfo {author} {\bibfnamefont {A.}~\bibnamefont
  {Blasco}}, \bibinfo {author} {\bibfnamefont {L.~J.}\ \bibnamefont {Garay}},
  \bibinfo {author} {\bibfnamefont {M.}~\bibnamefont {Mart\'in-Benito}}, \ and\
  \bibinfo {author} {\bibfnamefont {E.}~\bibnamefont
  {Mart\'{i}n-Mart\'{i}nez}},\ }\href {\doibase 10.1103/PhysRevLett.114.141103}
  {\bibfield  {journal} {\bibinfo  {journal} {Phys. Rev. Lett.}\ }\textbf
  {\bibinfo {volume} {114}},\ \bibinfo {pages} {141103} (\bibinfo {year}
  {2015})}\BibitemShut {NoStop}%
\end{thebibliography}%

\end{document}